\documentclass[prd,superscriptaddress,nofootinbib,twocolumn]{revtex4-1}
\usepackage{graphicx}
\usepackage{url}
\usepackage{color}
\usepackage{rotating}
\usepackage{amssymb,amsmath}

\def\mpcinv{{\rm Mpc}^{-1}}

\newcommand{\fnl}{f_{\rm NL}}

\newcommand{\zmin}{z_{\rm min}}
\newcommand{\zmax}{z_{\rm max}}

\newcommand{\kmsMpc}{\,{\rm km/s/Mpc}}
\newcommand{\fsky}{f_{\rm sky}}

\newcommand{\aap}{{A\&A}}
\newcommand{\apjs}{{Astrophys.~J.~Supp.}}
\newcommand{\aj}{{Astro.~Journal}}

\newcommand{\mnras}{{Mon.~Not.~R.~Astron.~Soc.}}

\newcommand{\Ylm}{Y_{\ell m}}
\newcommand{\tlm}{t_{\ell m}}
\newcommand{\clm}{c_{\ell m}}
\newcommand{\alm}{a_{\ell m}}
\newcommand{\wj}[6]{\left(
                           \begin{array}{ccc}
        \! #1\! & #2\!  & #3\!  \\
        \! #4\! & #5\!  & #6\!
                           \end{array}
                   \right)}
\newcommand*\mystrut[1]{\vrule width0pt height0pt depth#1\relax}

\newcommand{\Cl}{C_\ell}

\newcommand{\Clsys}{\mathcal{C}_{\ell_1}^{\rm sys}}

\newcommand{\ellmax}  {\ell_{\mathrm{max}}}
\newcommand{\ellcalibmax}  {\ell_{\mathrm{calib,max}}}

\newcommand{\Var}{\rm Var}
\newcommand{\FoM}{\rm FoM}
\newcommand{\nhat}{{\bf \hat n}}

\newcommand{\Nobs}{N_{\rm obs}}

\everymath{\displaystyle}

\begin{document}

\title{Calibration errors unleashed: effects on cosmological parameters and requirements for large-scale structure surveys}

\author{Dragan Huterer}
\affiliation{Department of Physics, University of Michigan, 
450 Church St, Ann Arbor, MI 48109-1040}

\author{Carlos Cunha}
\affiliation{Department of Physics, University of Michigan, 
450 Church St, Ann Arbor, MI 48109-1040}
\affiliation{Kavli Institute for Particle Astrophysics and Cosmology 452 Lomita Mall, Stanford University, Stanford, CA, 94305}

\author{Wenjuan Fang}
\affiliation{Department of Physics, University of Michigan, 
450 Church St, Ann Arbor, MI 48109-1040}
\affiliation{Department of Astronomy, University of Illinois at Urbana-Champaign, Urbana, IL 61801}

\date{\today}

\begin{abstract}
Imperfect photometric calibration of galaxy surveys due to either
astrophysical or instrumental effects leads to biases in measuring galaxy
clustering and in the resulting cosmological parameter measurements. More
interestingly (and disturbingly), the spatially varying calibration also
generically leads to violations of statistical isotropy of the galaxy
clustering signal. Here we develop, for the first time, a formalism to
propagate the effects of photometric calibration variations with arbitrary
spatial dependence across the sky to the observed power spectra and to the
cosmological parameter constraints. We develop an end-to-end pipeline to study
the effects of calibration, and illustrate our results using specific examples
including Galactic dust extinction and survey-dependent magnitude limits as a
function of zenith angle of the telescope.  We establish requirements on the
control of calibration so that it doesn't significantly bias constraints on
dark energy and primordial non-Gaussianity. Two principal findings are: 1)
largest-angle photometric calibration variations (dipole, quadrupole and a few
more modes, though not the monopole) are the most damaging, and 2) calibration
will need to be understood at the $\sim 0.1\%$--$1\%$ level (i.e.\ rms
variations mapped out to accuracy between 0.001 and 0.01 mag), though the precise
requirement strongly depends on the faint-end slope of the luminosity function
and the redshift distribution of galaxies in the survey.
\end{abstract}

\maketitle
\section{Introduction}\label{sec:intro}

Large-scale structure (LSS) measurements have become an extremely powerful probe of
cosmology over the past 30 years. Starting with the pioneering Harvard-CfA
survey \citep{Harvard_Cfa}, all the way to the Sloan Digital Sky Survey \citep{SDSS}
and its extension Baryon Oscillation Sky Survey \citep{BOSS},
Two-degree Field survey \citep{2dF}, and WiggleZ \cite{wigglez}, the LSS
surveys have revolutionized our understanding of the distribution of matter
and energy in the cosmos, and helped impose percent-level constraints on the
cosmological parameters (e.g. \cite{Anderson-BOSS}).

A major challenge in current and future imaging and spectroscopic LSS surveys
is understanding the sample selection.  We define calibration to be the
measure of our understanding of the selection of our sample of galaxies, and
calibration errors to be any unaccounted-for angular and redshift variations
in the selection.  The purpose of this paper is to determine how well
calibration errors need to be controlled in order to avoid substantial
degradation of the information we can extract from the LSS.

A particular source of uncertainty is known as {\it photometric calibration}.
The term refers to the adjustments required to establish a consistent spatial
and temporal measurement of flux of the target objects in the different bands
of observation throughout the entire photometric survey.  This is an enormous
problem that all existing and upcoming wide area surveys face.  The difficulty
comes from the variability of various building blocks of the observational
pipeline, which makes it difficult to establish a consistent flux baseline at
each band (i.e.\ the flux zeropoints).  In other words, because the instrument
sensitivity is constantly changing, and so are the sources and intensity of
noise, it is difficult to consistently compare the fluxes for objects at different
parts of the sky imaged at different times.  Some examples of the
manifestations of the photometric calibration errors in surveys are:
\begin{itemize}
\item {\it Detector sensitivity:} At any given time, the sensitivity of the
  pixels on the camera vary along the focal plane.  In addition, the
  sensitivity of a given pixel can change with time. 
\item {\it Observing conditions:} Wide area surveys are carried out over
  several years and conditions are constantly changing.  Observing conditions
  suffer from constant spatial and temporal variations.
\item {\it Bright objects:} The light from foreground bright stars and
  galaxies affects the sky subtraction procedure, which impairs the surveys'
  completeness near bright objects \citep{aih11,Ross}, and distorts the
  measured shapes of these faint galaxies \citep{man05}.
\item{\it Dust extinction:} Dust in the Milky Way absorbs light from the
  distant galaxies. As we show later, imperfect extinction correction can have
  serious consequences to cosmological clustering analyses.
\item{\it Star-galaxy separation:} In photometric surveys, faint stars can be
  erroneously included in the galaxy sample.  Conversely, galaxies are
  sometimes misclassified as stars and culled from the sample. These effects are
  important because stars are not randomly distributed across the sky.
\item{\it Deblending:} Galaxy images can overlap, and it can be difficult to
  cleanly separate photometric and spectroscopic measurements for the blended
  objects.
\end{itemize}
Variabilities in the instrument sensitivity and observing conditions cause an
angular variability in the depth of observations that the survey can achieve
through each filter.  Variations in the depth result in angular variations in
the number density and redshift distribution of objects.  In addition, because
galaxy spectra are not flat, and because the sample selection involves more
than one filter, depth variations cause variations in the angular and redshift
distribution of galaxy types.

This variability in the sample selection can, in principle, be accounted for.
This is not always done, however, and it is common, for example, for
correlation analyses of current data to assume a constant depth for the entire
survey.  Indeed, several sources of variability have been accounted for in the
analysis of existing data -- see in particular the pioneering work on the
subject in the modern era of LSS surveys by \citet{Scranton_sys} (see also
\citet{Vogeley_sys}), and the more recent efforts by \citet{Ho_sys} and
\citet{Ross}. These authors modeled a wide variety of systematic errors, some
of which qualify as the calibration errors (e.g.\ seeing, airmass, calibration
offsets).  In particular, the latter two papers identified bright stars as the
major contaminant which adds significant power to the intrinsic clustering
signal at large scales, and they applied two separate successful techniques to
subtract this systematic contamination.

For the upcoming surveys an even more detailed analysis will be needed,
ideally utilizing a formalism that is suited to a wide variety of photometric
calibration systematics mentioned above and captures any kind of
calibration-related systematic. One would also like to provide guidance on
how much calibration error, as a function of scale, can be tolerated in order
not to degrade the cosmological parameter inferences. Here we aim to address
both of these desiderata.

In this paper we set out to study calibration errors in the most general way
possible. Our goal is to build an end-to-end pipeline into which we can feed
calibration errors (or uncertainties) due to an arbitrary cause, and from
which we obtain biases in cosmological parameters inferred from measurements
of galaxy clustering in some LSS survey. We then turn the problem around, and
estimate how well the calibration errors need to be controlled in order not to
appreciably bias the cosmological parameter estimates.

To keep the scope of this paper reasonable, we only consider measurements of
the galaxy two-point correlation function (i.e.\ its Fourier transform, the
power spectrum), and leave other observable quantities -- higher-order
correlation functions of galaxies, for example -- for future work.  We also do
not consider the effect of the photometric redshift errors which, while very
important, are not expected to change our results in a major way, so we leave
the photo-zs for a future analysis.

The paper is organized as follows. in Sec.~\ref{sec:formalism} we describe our
formalism of modeling both the true, underlying galaxy density field and the
systematic errors describing variations in the photometric calibration.  In
Sec.~\ref{sec:biases} we present the formalism to derive cosmological
constraints and biases on cosmological parameters. In Sec.~\ref{sec:sens} we
propagate the effects of the systematic errors to calculate the biases in the
cosmological parameters. We conclude in Sec.~\ref{sec:concl}. Important 
technical details regarding various aspects of the computation of the effects
of the photometric variation systematics on the observable quantities are
relegated to the three Appendices.

\section{Formalism: describing spatially varying calibration}\label{sec:formalism}

In this section we start by defining calibration errors and their field
$c(\nhat)$, and proceed to derive the biased galaxy fluctuations in terms of
this field in multipole space. 

\subsection{Calibration errors: definition and basics}

Let true galaxy counts on the sky be denoted by $N(\hat{\bf n})$, where
$\hat{\bf n}$ is an arbitrary spatial direction. 
The survey mean is given by $\bar{N}\equiv \langle N(\hat{\bf
  n})\rangle_{\rm sky}$, where the average here is taken over the observed
sky. These true fluctuations in the galaxy counts can be expanded into 
harmonic coefficients $\alm$ as
\begin{equation}
\frac{N({\hat{\bf n}})-\bar{N}}{\bar{N}} =
\sum_{\ell=0}^\infty\sum_{m=-\ell}^\ell \alm\Ylm({\hat{\bf n}})
\label{eq:N_from_alm}
\end{equation}
Consider a survey where a {\it deterministic} calibration 
error $c({\hat{\bf n}})$ biases galaxy counts. In other words, given the true galaxy number
counts in some direction $N({\hat{\bf n}})$, the observed number is
\begin{equation}
\Nobs({\hat{\bf n}}) = [1+c({\hat{\bf n}})]\, N({\hat{\bf n}}), \label{eq:nobs}
\end{equation}
which implicitly defines the calibration field $c({\hat{\bf n}})$.
We can expand the calibration field relative to its fiducial value of zero
(corresponding to no error)
\begin{equation}
c({\hat{\bf n}}) =  \sum_{\ell_1=0}^{\ellcalibmax}\sum_{m_1=-\ell_1}^{\ell_1} 
c_{\ell_1 m_1}Y_{\ell_1 m_1}({\hat{\bf n}}),
\end{equation}
where hereafter we assume that the calibration error dominates on large
scales, and persists only out to some maximum multipole $\ellcalibmax$,
corresponding to the minimal angular scale of $\pi/\ellcalibmax$ radians.

The statistical properties of the two galaxy number-density field, and the
calibration-error field are, respectively
\begin{eqnarray}
\langle\alm\rangle &=& 0; \qquad \langle\alm a_{\ell'm'}^*\rangle = \delta_{mm'}\delta_{\ell
  \ell'} C_\ell\\[0.25cm]
\langle\clm\rangle &=& \clm; \quad \langle\clm c_{\ell'm'}^*\rangle = \clm c_{\ell'm'}^*
\end{eqnarray}
Throughout the paper, angular brackets $\langle\cdot\rangle$ indicate ensemble
averages, that is, averages over different realizations of the Universe. 
To reiterate, $N({\hat{\bf n}})$ is the Gaussian random, isotropic field as
predicted by inflation, while $c({\hat{\bf n}})$ is a deterministic function
given by calibration errors in the survey at hand.

In the remainder of this paper, we use the following definition: {\it
  Calibration variations (or errors) are departures of $c({\hat{\bf n}})$, or
  its harmonic coefficients $\clm$, from zero.} Our goal is to estimate how
accurately those variations have to be known in order not to bias the
cosmological parameter estimates.

Notice that we do not lose any generality by assuming that the c-field is
fixed, rather than stochastic like the true galaxy density field. In this
paper, we are effectively asking how much does this fixed systematic error
bias the usual cosmological constraints. We are, of course, free to iterate
over a number of specific incarnations of this 'fixed' error.  A more specific
example would be to ask how much does the Galactic dust pattern -- its
direction and amplitude fixed for the moment -- bias some cosmological
inference if unaccounted for perfectly, and then to repeat the analysis for a
number of dust pattern realizations, or even for several different dust models.

\subsection{Galaxy clustering and calibration errors: general case}

We now derive the main results regarding the effect of the
calibration errors on the observed clustering of galaxies.
Let us first calculate the observed density contrast of galaxies:
\begin{alignat}{2}
  \delta^{\rm obs}({\hat{\bf n}})  &\equiv&&  {\Nobs-\bar{N}_{\rm obs}\over\bar{N}_{\rm obs}}
  = \left [{N(1+ c({\hat{\bf n}}))\over \bar{N}(1+\epsilon)} - 1\right ]
  \nonumber \\[0.2cm]
  \label{eq:deltaobs}
  &=&& \frac{1}{1+\epsilon}\left [\delta ({\hat{\bf n}})\left (1+ c({\hat{\bf n}})\right ) + c({\hat{\bf n}}) -
    \epsilon \right ]\\[0.2cm]
  &=&& \frac{1}{1+\epsilon}\!\!\left [\delta ({\hat{\bf n}})
    \!\!\left (\!1\!+\! \sum_{\ell, m} \clm\Ylm\!\!\right )\! +\! \sum_{\ell, m} \clm\Ylm \!-\!
    \epsilon \right ]\nonumber 
\end{alignat}
where $\Ylm\equiv \Ylm ({\hat{\bf n}})$, and we expanded the photometric
calibration variation field $c({\hat{\bf n}})$ into spherical harmonics. Here
$\epsilon$ is the relative bias in the measured mean number  of galaxies:
\begin{eqnarray}
\bar{N}_{\rm obs} &\equiv & \langle \Nobs({\hat{\bf n}}) \rangle_{\rm sky}    \nonumber\\[0.2cm]
&=& \bar{N} + \sum_{\ell m} \clm\langle N({\hat{\bf n}})\Ylm({\hat{\bf
    n}})\rangle_{\rm sky} \label{eq:Nobs_over_N_eps}\\[0.2cm]
&\equiv & \bar{N}(1+\epsilon)\nonumber
\end{eqnarray}
(the $\langle\rangle_{\rm sky}$ denotes sky average), so that 
\begin{equation}
\epsilon\equiv \frac{1}{\bar{N}}\sum_{\ell m}\clm\langle N({\hat{\bf n}})\Ylm({\hat{\bf
    n}})\rangle_{\rm sky} .
\label{eq:epsilon_app}
\end{equation}
The quantity $\epsilon$ can be evaluated directly in real space as above when given the
calibration error map, or in harmonic space, combining Eq.~(\ref{eq:epsilon_app}) and
Eq.~(\ref{eq:N_from_alm})
\begin{equation}
\epsilon = {c_{00}\over \sqrt{4\pi}} + \sum_{\ell, m}{\clm\alm^*\over 4\pi}
\label{eq:eps_in_terms_of_clm}
\end{equation}
where we used the identity $(-1)^ma_{\ell (-m)} = \alm^*$ and the
orthogonality relation for  spherical harmonics. In cases where $\fsky<1$, the
orthogonality relation does not hold, but Eq.~(\ref{eq:eps_in_terms_of_clm})
still does if the coefficients $\alm$ are interpreted as the cut-sky harmonics
of the density field.  

The observed galaxy overdensity field can also be expanded in terms of the harmonic
basis
\begin{equation}
\delta^{\rm obs}(\hat{\bf n}) 
\equiv t(\hat{\bf n}) = \sum_{\ell m}\tlm\Ylm({\hat{\bf n}}),
\end{equation}
Equating this to the last expression in Eq.~(\ref{eq:deltaobs}) and inverting by
multiplying with $\Ylm^*$ and using the orthogonality relation, we obtain the 
harmonic coefficients of the observed galaxy overdensity field $\tlm$ in terms the
true galaxy fluctuation field $\alm$ and the calibration field $\clm$
\begin{eqnarray}
\tlm  &=& \frac{1}{1+\epsilon}\left [\alm + \clm + \!\!\sum_{\ell_1 \ell_2 m_1 m_2}\!\!\!
R_{m_1 m_2 m}^{\ell_1 \ell_2 \ell}\, c_{\ell_1 m_1} a_{\ell_2 m_2} \right. \nonumber  \\[0.2cm] 
&-& \left . \sqrt{4\pi}\epsilon\,\delta_{\ell 0}\,\delta_{m0}\right ]
\label{eq:tlm}
\end{eqnarray}
where to obtain the last term in the last line we used $1=\sqrt{4\pi}Y_{00}$.
Here we define the coupling matrix $R$ in terms of Wigner 3j
symbols
\begin{eqnarray}
R_{m_1 m_2 m}^{\ell_1 \ell_2 \ell} &\equiv & 
(-1)^m \sqrt{ (2\ell_1+1)(2\ell_2+1)(2\ell+1) \over 4\pi}  \nonumber \\[0.15cm]
&&\times \wj{\ell_1}{\ell_2}{\ell}{0}{0}{0} 
\wj{\ell_1}{\ell_2}{\ell}{m_1}{m_2}{-m} \,.
\label{eq:R}
\end{eqnarray}

Calculating the two-point correlation of $\tlm$ is now straightforward, and things
are simplified because all terms proportional to a single power of $\alm$ (or
its conjugate) vanish -- recall that $\clm$ are just some numbers
here. Moreover, we can ignore the term proportional to $\delta_{\ell 0}$ --
last term in Eq.~(\ref{eq:tlm}) -- since it only affects the monopole which
is not used in cosmological constraints.  The ensemble average of the
multipole moments becomes, after some algebra
\begin{widetext}
\begin{equation}
\langle  t_{\ell m} t_{\ell' m'}^*\rangle  =
\frac{1}{(1+\epsilon)^2}  \left \{
\underbrace{\mystrut{3.0ex}\delta_{mm'}\delta_{\ell \ell'} C_\ell}_{\rm isotropic}+
\underbrace{\left[U^{\ell' \ell}_{m' m}\,C_{\ell'} + (U^{\ell
    \ell'}_{mm'})^*\, C_\ell\right ]
+\sum_{\ell_2 m_2} U^{\ell_2 \ell}_{m_2 m}(U^{\ell_2 \ell'}_{m_2m'})^*\,
C_{\ell_2} + \clm c_{\ell'm'}^*}_{\rm breaks\,\, statistical\,\, isotropy}
\right \}
\label{eq:tlm_tl'm'}
\end{equation}
\end{widetext}
where we defined 
\begin{equation}
U^{\ell_2 \ell}_{m_2 m}\equiv \sum_{\ell_1 m_1}c_{\ell_1 m_1}R_{m_1 m_2
  m}^{\ell_1 \ell_2 \ell}
\label{eq:U_in_terms_of_R}
\end{equation}
which is a function that depends on the Wigner 3j symbols as well as the
calibration-field coefficients $c_{\ell m}$. 

Equation (\ref{eq:tlm_tl'm'}) is the key result in this paper. As the label in
the equation shows, the observed galaxy density field $t(\hat{\bf n})$
exhibits {\it broken statistical isotropy}. In particular, the variance of $t$
is not rotationally invariant any more (i.e.\ it depends on $m$), and
covariance between the different $\ell$ modes is not zero any more.  We can
now utilize this formula and consider the isotropically measured power
(i.e.\ assuming $\ell=\ell'$ and averaging over $m=m'$) and estimate how
accurately any given systematic, described by the full set of $c_{\ell m}$,
needs to be understood in order not to degrade the accuracy in measuring the
cosmological parameters including non-Gaussianity\footnote{One could also
  study how well one can utilize the full power of LSS measurements -- by {\it
    not} assuming statistical isotropy (i.e.\ the full $\ell\ell'mm'$
  dependent expression) -- to detect, and potentially correct for, the
  systematics. We will study prospects for such ``self-calibration'' --
  determination of the systematic errors internally from the survey, utilizing
  the $\ell\neq\ell'$, $m\neq m'$ correlators -- in a future work.}.

\subsection{Galaxy clustering and calibration errors: isotropic power case}\label{sec:bias_isotropic}

We usually -- essentially always, in fact! -- assume that the field is
isotropic, and then we use the data to calculate the correlation function,
power spectrum, etc. Let us see how the assumed-isotropic angular power
spectrum is biased in terms of an arbitrary contamination field. 

Setting $\ell=\ell'$ and $m=m'$ in Eq.~(\ref{eq:tlm_tl'm'}), we get
%
\begin{eqnarray}
\langle |\tlm |^2\rangle  &=&
\frac{1}{(1+\epsilon)^2} \left (
C_\ell +
2\left(U^{\ell \ell}_{m m}\right )^{\rm Re}\,C_{\ell} \right . \nonumber \\[0.2cm]
&+& \left . \sum_{\ell_2 m_2} \left |U^{\ell_2 \ell}_{m_2 m}\right |^2\, C_{\ell_2} 
+ |\clm |^2\right ). 
\label{eq:tlm_tlm}
\end{eqnarray}
To assume statistical isotropy, we not only set $\ell=\ell'$ and $m=m'$ but
further average over the $2\ell+1$ values of $m$ for a fixed $\ell$. Then we
obtain the prediction for the angular power spectrum that one would measure
{\it assuming} statistical isotropy even when the systematics break it:
\begin{widetext}
\begin{equation}
T_\ell \equiv  \frac{\sum_{m=-\ell}^\ell \langle|\tlm  |^2\rangle}{2\ell+1} =
\frac{1}{(1+\epsilon)^2} \left [
\left (1+2\, \frac{c_{00}}{\sqrt{4\pi}}\right )C_\ell
+\frac{\sum_{m=-\ell}^\ell \left (
\sum_{\ell_2 m_2} |U^{\ell_2 \ell}_{m_2 m}|^2\, C_{\ell_2} +|\clm|^2\right )}
{2\ell+1}
\right ]
\label{eq:Tl}
\end{equation}
\end{widetext}
where $|U|^2\equiv UU^*$, and where the sum over $\ell_2$ goes in principle
over all multipoles (though only those from the range $[\ell-\ellcalibmax,
  \ell+\ellcalibmax]$ are nonzero), while $m_2$ goes from $-\ell_2$ to
$\ell_2$. Note that the term linear in $U$ seen in Eq.~(\ref{eq:tlm_tlm})
dramatically simplified in the expression for $T_\ell$ (Eq.~\ref{eq:Tl}) after
we used the summation relation
\begin{equation}
\sum_{m=-\ell}^\ell (-1)^m\, \wj{\ell_1}{\ell}{\ell}{0}{m}{-m} \,
= (-1)^\ell \sqrt{2\ell+1}\,\delta_{\ell_1 0}.
\end{equation}

For a pure monopole calibration error (i.e.\ a pure $c_{00}$ term), one can
verify that the effects of the $\epsilon$ term and the $c_{00}$ term in
Eq.~(\ref{eq:Tl}) exactly cancel and $T_\ell$ is unchanged. This makes
intuitive sense, as a shift in the monopole changes the mean counts on the
sky but does not affect the density {\it fluctuations}.

One can intuitively understand the individual terms on the right-hand side of  Eq.~(\ref{eq:Tl}):
\begin{itemize}
\item The $(1+\epsilon)^{-2}$ prefactor accounts for the change in the mean number of
  galaxies observed on the sky; see Eq.~(\ref{eq:Nobs_over_N_eps}). As already
  mentioned, this term effectively ensures that the effect of the monopole
  ($c_{00}$, or constant change in calibration across the sky) is precisely canceled out.
\item The terms containing $U^{\ell_2 \ell}_{m_2 m}$ introduce coupling
  between the multipoles. In particular, if there are calibration error
  multipoles out to some multipole $\ellcalibmax$, then the galaxy power at
  multipole $\ell$ will be contaminated by contributions coming from the
  range $[\ell-\ellcalibmax, \ell+\ellcalibmax]$, a fact familiar from spin
  coupling in quantum mechanics.
\item Presence of the term $|\clm|^2$ essentially means that the power of the
  calibration field is added to the intrinsic galaxy power spectrum. In other
  words, even if the distribution of galaxies on the sky were perfectly
  uniform so that $\delta\equiv N/\bar{N}-1=0$ over some area, the calibration
  field will induce power so that $\delta^{\rm obs}\neq 0$.
\end{itemize}

From the structure of Eq.~(\ref{eq:Tl}), it is clear that calculating and
storing the coefficients $U$ is challenging.  Naively, the problem requires
evaluation of roughly $10^{18}$ coefficients.  Appendix \ref{app:tabulate}
describes our approach of limiting the number of evaluations of $(\ell,
\ell_1, \ell_2, m, m_1, m_2)$ and tabulating the coefficients $U$ so that the
number of operations is only of order $10^8$ (for $\ellmax=1000$ binned in
$\sim 30$ multipole bins and considering the calibration variations out to
$\ellcalibmax=20$), and is thus feasible.  We plot the biased power spectra
$T_\ell$ further below in the next section.

Finally, it is worth writing down the observed angular {\it cross-correlation}
power spectrum between fluctuations $\tlm^{(i)}$ and $\tlm^{*(j)}$ in two
different tomographic redshift bins $i$ and $j$; it follows straightforwardly
from Eq.~(\ref{eq:Tl}) that:
\begin{widetext}
\begin{equation}
T_{\ell}^{(ij)} =
\frac{1}{(1+\epsilon^{(i)})(1+\epsilon^{(j)})} \left [
\left (1+\, \frac{c_{00}^{(i)}+c_{00}^{(j)}}{\sqrt{4\pi}}\right )C_{\ell}^{(ij)}
+\frac{\sum_{m} \left (
\sum_{\ell_2 m_2} (U^{\ell_2 \ell}_{m_2 m})^{(i)}\,(U^{\ell_2 \ell}_{m_2 m})^{*(j)}\, C_{\ell_2}^{(ij)} +\clm^{(i)}\clm^{*(j)}\right )}
{2\ell+1}
\right ]
\label{eq:Tl_tomo_bins}
\end{equation}
\end{widetext}
where $\clm^{(i)}$, $\epsilon^{(i)}$ and $(U^{\ell_2 \ell}_{m_2 m})^{(i)}$ are
all evaluated in the redshift bin $i$ (and same for $j$), and where
  $C_\ell^{(ij)}$ are the true galaxy cross-correlation power spectra. 
While physical sources of calibration error are typically local and thus
redshift-independent, in Sec.~\ref{sec:sens} we demonstrate that converting
from the magnitude error to the calibration field $c(\hat{\bf n})\equiv
(\delta N/N)(\hat{\bf n})$ depends on the faint-end slope of the luminosity
function, which typically {\it is} redshift-dependent, hence making the
harmonic coefficients of $c(\hat{\bf n})$ also z-dependent and thereby
potentially introducing couplings between the different redshift bins.

\subsection{Additive and multiplicative systematics}

Before we obtain numerical results on how some realistic calibration errors
affect the observed power spectra, it pays to consider qualitatively how the
angular power spectrum of galaxies is affected.

It is often useful to divide the effect of systematic errors into {\it
  additive} (those whose field is added to the true field observed on the
sky), and {\it multiplicative} (those whose field multiplies the true field);
see e.g.\ \cite{Gordon2005} where his nomenclature has been previously
employed in the cosmic microwave background (CMB) context, and \cite{HTBJ} and
\cite{Heymans_STEP1} who considered the additive and multiplicative systematic
errors in weak lensing measurements.  These terms refer to the systematic
error that either adds to the true galaxy fluctuations, or else multiplies it
and modulates the true signal; see Eq.~(\ref{eq:deltaobs}) for the real-space
and Eq.~(\ref{eq:tlm}) for the harmonic-space picture.  For example, on the
right-hand side of Eq.~(\ref{eq:tlm}) the term $\alm$ corresponds to the true
density field, $\clm$ represents the additive effect of the systematic error,
while the term containing $\clm\alm$ term together with the geometric factor
$R$ and the appropriate sum represents the multiplicative effect of the
systematics\footnote{Technically speaking, the systematic effects are all
  multiplicative in Eq.~(\ref{eq:tlm}) because of the $1/(1+\epsilon)$
  prefactor; however given that this prefactor is typically very close to unity,
  the second and third term in parentheses of this equation act approximately
  in the additive and multiplicative sense.}.

Additive and multiplicative errors in the counts translate into additive and
multiplicative contributions from the calibration field to the observed galaxy
power spectrum, see Eq.~(\ref{eq:tlm_tl'm'}): the additive error is the term
$\clm\clm^*$ while the multiplicative error are
all terms involving the coupling matrix $U$.  The two kinds of errors produce
qualitatively different effects: additive error at some multipole $\ell_1$
only affects power at that multipole, while the multiplicative error affects
power at a range of multipoles; in particular, true power at an arbitrary
multipole $\ell$ would leak to all multipoles in the range $[\ell-\ell_1,
  \ell+\ell_1]$.

The additive terms dominate the error budget on the largest scales, but are
subdominant at smaller scales \citep{Vogeley_sys}. This can be understood
qualitatively as follows: modulo geometric coupling terms, both additive and
multiplicative terms are proportional to the square of coefficients $\clm$,
but the multiplicative terms are further multiplied by the fiducial angular
power $\Cl$. Given that $\Cl\ll 1$ at all $\ell$ and any redshift\footnote{One
  exception -- $\Cl$ of order unity or larger -- is realized in the scenario
  of galaxies at a very low redshift, $z\lesssim 0.05$. However such a sample
  would probably not be useful for cosmology given the significant galaxy
  peculiar velocities and the necessarily small volume probed. Moreover, the
  calibration errors would have less impact to begin with, since they would be
  affecting a very large intrinsic clustering signal.}, the multiplicative
terms are suppressed relative to the additive terms. At higher multipoles, on
the other hand, there are more ways in which power from other scales can leak
into that $\ell$ so that the sums associated with multiplicative terms make
them the dominant systematic contribution.

While it is often assumed for simplicity that the systematic errors,
calibration or other, are represented by purely additive errors, we just
demonstrated that both kinds of errors are important. In fact, in the
plausible scenario where largest-scale information in the survey is ignored
to avoid the systematic contamination, the multiplicative errors dominate. In
what follows we use the full expressions containing both additive and
multiplicative terms. 

\section{power spectra and their biases}\label{sec:biases}

In this section we propagate the effect of calibration errors to estimate
biases in the cosmological parameters describing dark energy and primordial
non-Gaussianity. 

\subsection{Fiducial cosmological model}

We consider a set of cosmological parameters with the following fiducial
values: matter density relative to critical $\Omega_M=0.25$, dark energy equation of state
parameter today $w_0=-1$, its variation with scale factor $w_a=0$, spectral
index $n=0.96$, and amplitude of the matter power spectrum $\ln A$ where
$A=2.3\times 10^{-9}$ (corresponding to $\sigma_8=0.80$) defined at scale
scale $k=0.002 \mpcinv$. Note that we hold fixed the Hubble constant
$h=H_0/(100 \kmsMpc)$ (or equivalently, physical matter density $\Omega_M
h^2$), the physical baryon density $\Omega_B h^2$, and we assumed a
  flat universe. On the other hand, we do not assume any other prior
information, such as the CMB information from WMAP and Planck. In practice,
this prior information would largely serve to fix $h$, $\Omega_M h^2$, $\Omega_B h^2$, and 
  curvature $\Omega_K$. Note that this rather restricted set of assumptions about the set
of cosmological parameters and external information about them is sufficient
for our analysis: we are primarily concerned about the effect of the
calibration systematics on the measured power spectra, and on the biases in
the dark energy and non-Gaussianity parameters. Given that the systematics
strongly depend on the properties of the galaxy sample and the survey (as we
discuss further below), it is not necessary to model the up-to-date knowledge
about the cosmological parameters in great detail.

\begin{figure*}
\centering
\includegraphics[width=0.45\linewidth]{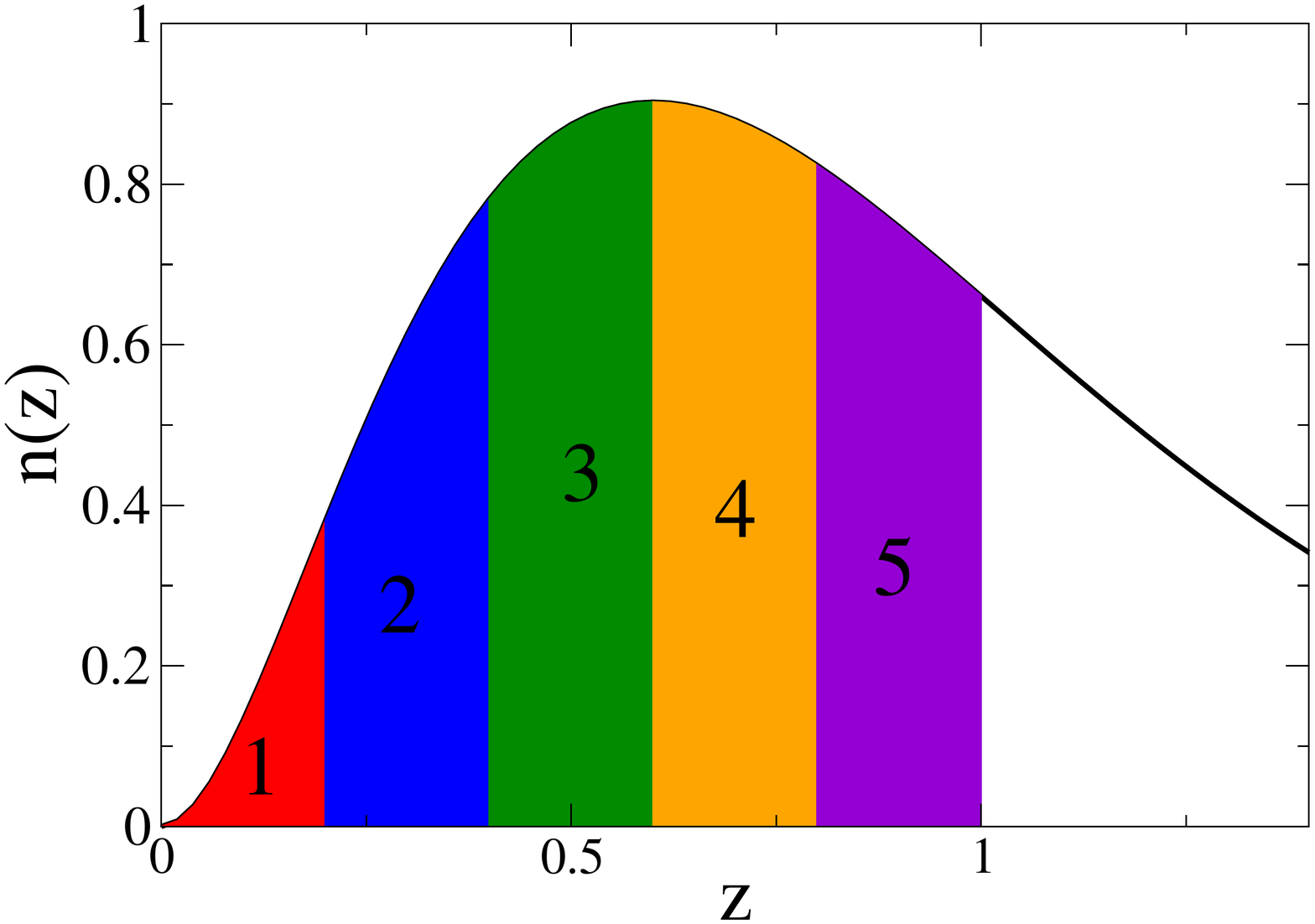} 
\includegraphics[width=0.45\linewidth]{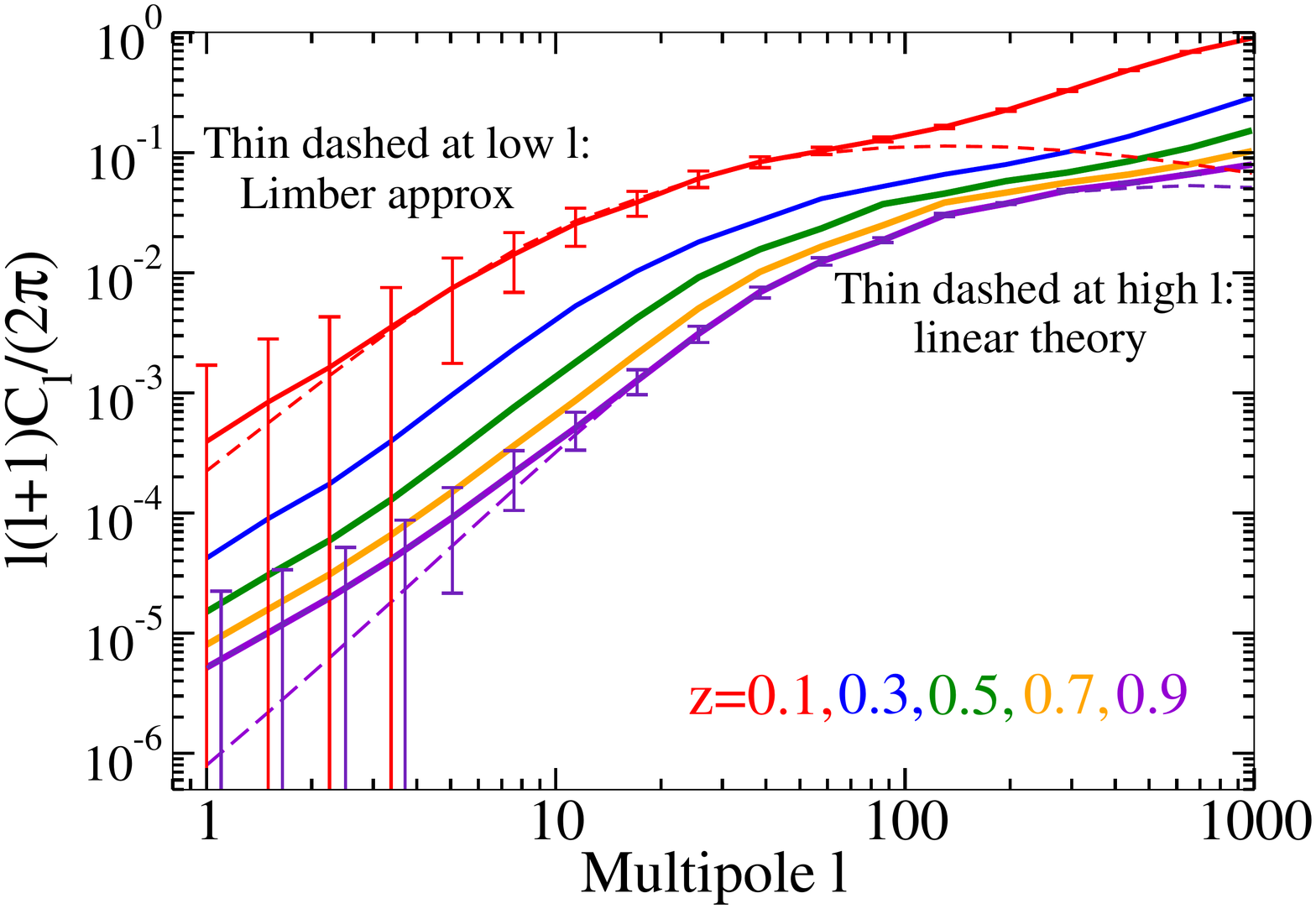} 
\caption{Left panel: density distribution of galaxies assumed in this paper,
  and boundaries of the five redshift bins. We ignore information at $z>1$,
  thus roughly modeling the difficulties with establishing accurate
  photometric redshifts at that range (for the DES). Right panel: Angular
  power spectra $C_\ell^{(ii)}$ for five redshift bins [Cross-correlations
    between the bins, while used in the analysis, are very small and not important
    nor shown in the figure.]. For the first and fifth bin we show, at the low
  multipole end, the full expression that we use at large scales (see
  Eq.~(\ref{eq:noLimber})) and, at the high-multipole end, the linear power
  spectrum for reference. For the first and fifth redshift bin we also show
  the cosmic variance errors plus shot noise. }
\label{fig:Cl}
\end{figure*}

We assume a survey covering 5,000 square degrees (so $\fsky\simeq 0.12$) with
information out to $\zmax=1$, corresponding roughly to the Dark Energy Survey
(DES). We assume that the number density of the galaxies is $n(z)\propto
z^2\exp(-z/z_0)$ with $z_0=0.3$, and that photometric redshifts enable
splitting the sample in five tomographic bins centered at $z=0.1$, 0.3, 0.5,
0.7 and $0.9$; see the left panel of Fig.~\ref{fig:Cl}. The fiducial
statistical constraints on the dark energy parameters are $\sigma(w_0)=0.06$
and $\sigma(w_a)=0.24$.

Instead of the power spectrum in wavenumber $P(k)$, we consider measurements
of the angular power spectrum of galaxy fluctuations. In the Limber
approximation, which is valid on intermediate to small angular scales, the
angular power is given as (e.g.\ \citep{EDSGC,Zhan_Knox_Tyson})
\begin{eqnarray}
\label{eq:Limber}
P_{\ell}^{(ij)} &\equiv& \langle   \alm^{(i)}\alm^{*(j)}\rangle \\[0.2cm]
&=& {2\pi^2\over \ell^3}\!\! \int_0^\infty \!\!
r(z) H(z) b^2(k, z) \Delta^2(k, z) W_i(z) W_j(z)dz
\nonumber
\end{eqnarray}
where $i$ and $j$ are referring to one of the five redshift
bins\footnote{While we use the all of the cross-correlations, $i\gtrsim j$, we
  notice that the cosmological constraints -- and requirements on the
  systematics -- remain unchanged if we only use the angular power spectrum
  auto-correlations with $i=j$. This is not too surprising, as
  cross-correlations $\Cl^{(ij)}$ are guaranteed to be zero in the Limber
  approximation and in the absence of photometric redshift errors (photo-zs
  generically lead to overlap of redshift bins and thus bin-to-bin
  correlations). At large scales, where we do not employ the Limber
  approximation because it is not accurate there (especially for nonzero $\fnl$ which
  introduces very large-scale correlations), we explicitly verify that
  cross-powers do not appreciably alter our requirements on the control of the
  calibration errors. } and $b(k, z)$ is the bias defined below.  The full,
beyond-Limber expression, as well as the definition of the $W(z)$ terms, are
given in Appendix \ref{app:PS}. The power spectrum $\Delta^2(k, z)\equiv k^3
P(k)/(2\pi^2)$ is calculated using the transfer function output by CAMB, and
its nonlinearities are modeled with the \citet{smith07a} formulae that were
based on a halo model and fit to simulations. Appendix \ref{app:PS} has all of
the details of how we calculate the power spectrum.

We consider information from multipoles $1\leq \ell\leq 1000$, corresponding
to spatial scales from about 10 arcmin to 180 degrees.
To obtain accurate constraints on the parameter $\fnl$ which come from large
angular scales,  we use every individual multipole between $\ell=1$
and $\ellcalibmax=20$; beyond this we use ten more widely separated bins with $\Delta
\ell\simeq 100$. Therefore, we use a total of 30 bins in $\ell$; for the
low-$\ell$ ones we do {\it not} assume the Limber approximation and use
Eqs.~(\ref{eq:noLimber}), while for the higher multipoles we use the Limber
approximation, Eq.~(\ref{eq:Limber}) above.

Finally, we also allow for the presence of primordial non-Gaussianity. We adopt
the widely studied `local' model of non-Gaussianity
\begin{equation}
\Phi({\bf x}) = \Phi_G({\bf x}) +\fnl (\Phi_G^2-\langle\Phi_G^2\rangle),
\label{eq:fnl}
\end{equation}
(where $\Phi$ is the primordial Newtonian gravitational potential, $\Phi_G$ is
its Gaussian component, and $\fnl$ is a dimensionless parameter), the bias
becomes scale-dependent, with a new term that goes as $k^{-2}$ \cite{Dalal}
\begin{equation}
b(k)=b_0 + \fnl(b_0-1)\delta_c\, \frac{3\Omega_MH_0^2}{a\,g(a) T(k)c^2 k^2},
\label{eq:bias}
\end{equation}
where $b_0$ is the usual Gaussian bias (on large scales, where it is
constant), $\delta_c\approx 1.686$ is the collapse threshold, $a$ is the scale
factor, $\Omega_M$ is the matter density relative to critical, $H_0$ is the
Hubble constant, $k$ is the wavenumber, $T(k)$ is the transfer function,
$g(a)$ is the growth suppression factor, and $c$ is the speed of
light\footnote{This formula has higher-order corrections calibrated to N-body
  simulations and also derivable from theory, but these are typically small
  and not crucial for the present analysis so we ignore them here.}. We assume
the fiducial model with the Gaussian bias $b_0=2$ and zero non-Gaussianity, $\fnl=0$.

The full set of cosmological parameters that we use is therefore
\begin{equation}
p_a \in \{\Omega_M, w_0, w_a, n, A, \fnl \}.
\label{eq:params}
\end{equation} 
The cosmological constraints can then be computed from the Fisher matrix
\begin{equation}
F_{ab} = 
\sum_{\ell, \alpha, \beta}
\,{\partial C_{\ell}^{(\alpha)} \over \partial p_a}
\,{\rm Cov}^{-1}\left [C_{\ell}^{(\alpha)}, 
  C_{\ell}^{( \beta)}\right ]
\,{\partial C_{\ell}^{(\beta)} \over \partial p_b},
\label{eq:Fisher}
\end{equation}
where $\alpha$ and $\beta$ stand for all pairs of bin indices $(i, j)$ with
$i\leq j$ (since $C_{\ell}^{(ji)}\equiv C_{\ell}^{(ij)}$). The observed power spectrum is
equal to the raw power plus shot noise
\begin{equation}
C_{\ell}^{(ij)} = P_{\ell}^{(ij)} + \delta_{ij}{1 \over N^{\rm sr}_i}
\end{equation}
where $N^{\rm sr}_i$ is the number of galaxies per steradian in the
tomographic bin $i$.  Moreover, ${\bf Cov}^{-1}$ in Eq.~(\ref{eq:Fisher}) is
the inverse of the covariance matrix between the observed power spectra;
assuming observations in the linear (and therefore Gaussian) regime only, the
covariance matrix follows directly from Wick's
theorem:
\begin{eqnarray}
{\rm Cov}\left [C_{\ell}^{(ij)}, C_{\ell'}^{(kl)}\right ] &=& 
{\delta_{\ell\ell'}\over (2\ell+1)\,f_{\rm sky}\,\Delta \ell}\,
\\[0.2cm]
&\times &\left [ C_{\ell}^{(ik)} C_{\ell}^{(jl)} + 
  C_{\ell}^{(il)} C_{\ell}^{(jk)}\right ].\nonumber
\label{eq:Cov}
\end{eqnarray}
The minimal error in the $i$-th cosmological parameter is,
by the Cram\'er-Rao inequality, $\sigma(p_i)\simeq \sqrt{(F^{-1})_{ii}}$.

Finally, we would like to estimate the bias in the cosmological parameters,
$\delta p_a$, given an arbitrary systematic error in the power-spectrum,
$\delta C_\ell$. The bias can be estimated using the Fisher matrix formalism
as follows:
\begin{equation}
\!\delta p_a = \sum_b
F_{ab}^{-1} \sum_{\ell, \alpha, \beta}
\delta C_{\ell}^{(\alpha)}
\,{\rm Cov}^{-1}\left [C_{\ell}^{(\alpha)}, 
  C_{\ell}^{( \beta)}\right ]
\,{\partial C_{\ell}^{(\beta)} \over \partial p_b}.
\label{eq:deltap}
\end{equation}

\subsection{Biases in the observed power spectra}

We would like to fairly compare the biases as a function of $\ell_1$, so we
choose to adopt coefficients describing the uncertainties in the calibration
field $c_{\ell_1 m_1}$ that lead to a {\it fixed variance} in the calibration
pattern on the sky $c(\hat{\bf n})$ (and corresponding to, as we will shortly
see, fixed variance in the angular variations of the magnitude limits of the
survey).  Thus, we have
\begin{equation}
\Var(c(\hat{\bf n})) = {2\ell_1+1\over 4\pi}\Clsys\\[0.2cm]
= {|c_{\ell_1 m_1}|^2\over 4\pi} \quad \mbox{(no sum)}
\end{equation}
since $\Clsys=|c_{\ell_1 m_1}|^2/( 2\ell_1+1)$ is the angular power spectrum
of the systematics (and really just the sum of their coefficients squared) and
where the reader is reminded that, in this particular calculation we are
``turning on'' one $(\ell_1, m_1)$ pair at a time.

To consider the calibration variation in a single multipole ($\ell_1$, $m_1$)
that leads to a fixed variance in $c(\hat{\bf n})$, we therefore make the
choice
\begin{equation}
c_{\ell_1 m_1}^{\rm Re, Im} = 
\left \{  \begin{array}{cl}
\sqrt{4\pi\Var(c(\hat{\bf n}))}  \quad (m=0) \\[0.4cm]
\sqrt{2\pi\Var(c(\hat{\bf n}))}  \quad (m\neq 0) \\[0.2cm]
\end{array} \right .
\label{eq:clm_var_const}
\end{equation}
where in the $m\neq 0$ case both real and imaginary part of the $c_{\ell_1
  m_1}$ have the given value.

\begin{figure}[t]
\centering
\includegraphics[width=\linewidth]{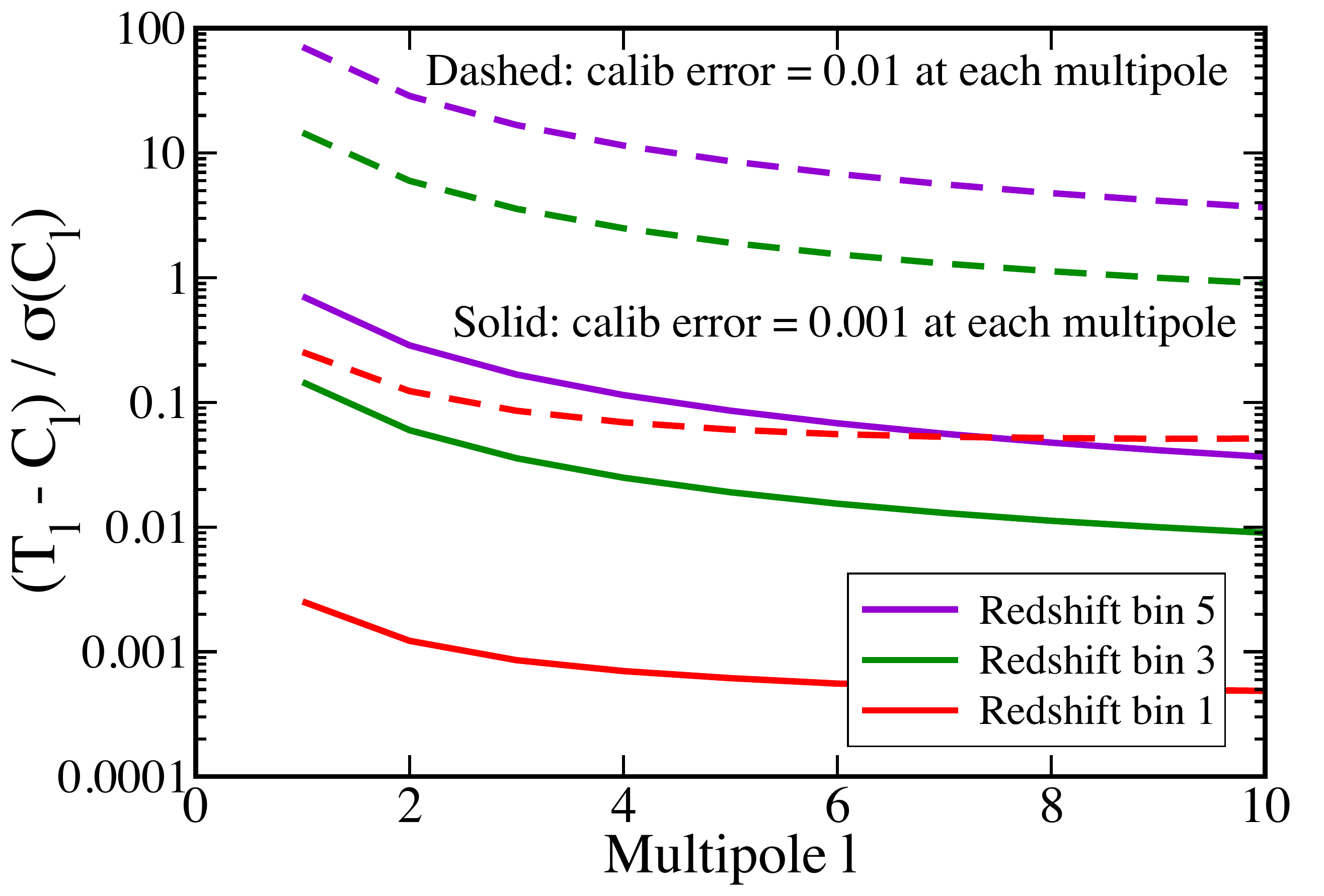} 
\caption{Difference between the the observed isotropic part of the power
  spectrum $T_\ell$ (see Eq.~(\ref{eq:Tl})) and the fiducial $C_\ell$, divided by
  the statistical error (cosmic variance plus shot noise) for our assumed
  DES-type survey with $\fsky\simeq 1/8$. We assume a constant rms photometric
  variation of 0.01 (dashed curves) or 0.001 (solid curves) per each multipole $\ell$. Note
  that higher redshift bins are affected more than the lower redshifts. 
  The fall-off with $\ell$ can be understood
  analytically; see text for details.}
\label{fig:Tl_minus_Cl_over_sigma}
\end{figure}

Figure \ref{fig:Tl_minus_Cl_over_sigma} shows the difference between the
observed isotropic part of the power spectrum $T_\ell$ (see Eq.~(\ref{eq:Tl}))
and the fiducial $C_\ell$, divided by the statistical error (cosmic variance plus
shot noise) for our assumed DES-type survey with $\fsky\simeq 1/8$. We assume
a constant calibration error with rms\footnote{For the
    purposes of making Figure \ref{fig:Tl_minus_Cl_over_sigma} we put all of
  the photometric calibration variation in a single value of
  $m_1=0$. } 
of 0.01 or 0.001 per each multipole $\ell_1$
separately. While a fixed $\ell_1$ of the systematic errors affects all
multipoles $\ell$, it affects $\ell=\ell_1$ the most, so in this graph we only
plot the effect on the observed power spectrum at the {\it same multipole} at
which the systematic errors occurs. In other words, Figure
\ref{fig:Tl_minus_Cl_over_sigma} shows the maximally affected multipole
$\ell$, for a fixed calibration variation error.
The error in the measured tomographic angular power spectra decreases with
multipole $\ell$, which can be understood easily as follows: to a good
approximation, the additional terms in the observed isotropic power spectrum,
Eq.~(\ref{eq:Tl}), are dominated by the additive term $|\clm|^2/(2\ell+1)$. We
find this is the case even for the lowest-redshift tomographic bin where $\Cl$
is the highest and thus helps boost the multiplicative terms that contain the
$U$ coefficients.  Moreover, generally we find that $\epsilon\simeq 0$.
Therefore,
\begin{equation}
{T_\ell-\Cl\over \sigma(\Cl)} \simeq {\displaystyle{|\clm|^2\over (2\ell+1)}\over 
\displaystyle\sqrt{2\over (2\ell+1)\fsky }\,
\left (\Cl + \frac{1}{N^{\rm sr}}\right )}\label{eq:tl-cl}
\end{equation}
where $\Cl\equiv C_\ell^{(ii)}$ is referring to the power spectrum
  in some redshift bin $i$, and same for $T_\ell$ and $\clm$. For a fixed variance in the photometric
  variations, $\clm$ is independent of $\ell$, so that this expression goes as
  $1/[\sqrt{2\ell+1}\Cl]$, decreasing with $\ell$ approximately
  as $\ell^{-1.3}$, at least out to $\ell=10$ plotted in this Figure.

Figure \ref{fig:Tl_minus_Cl_over_sigma} further shows that higher redshift
bins are affected more than the lower redshifts.  This is easy to understand:
at higher redshift, the cosmological signal is smaller, given that it averages
over more large-scale structure along the line of sight, and therefore it is
more susceptible to the (redshift-independent) calibration bias.

\begin{figure*}[t]
\centering
\includegraphics[scale=0.5]{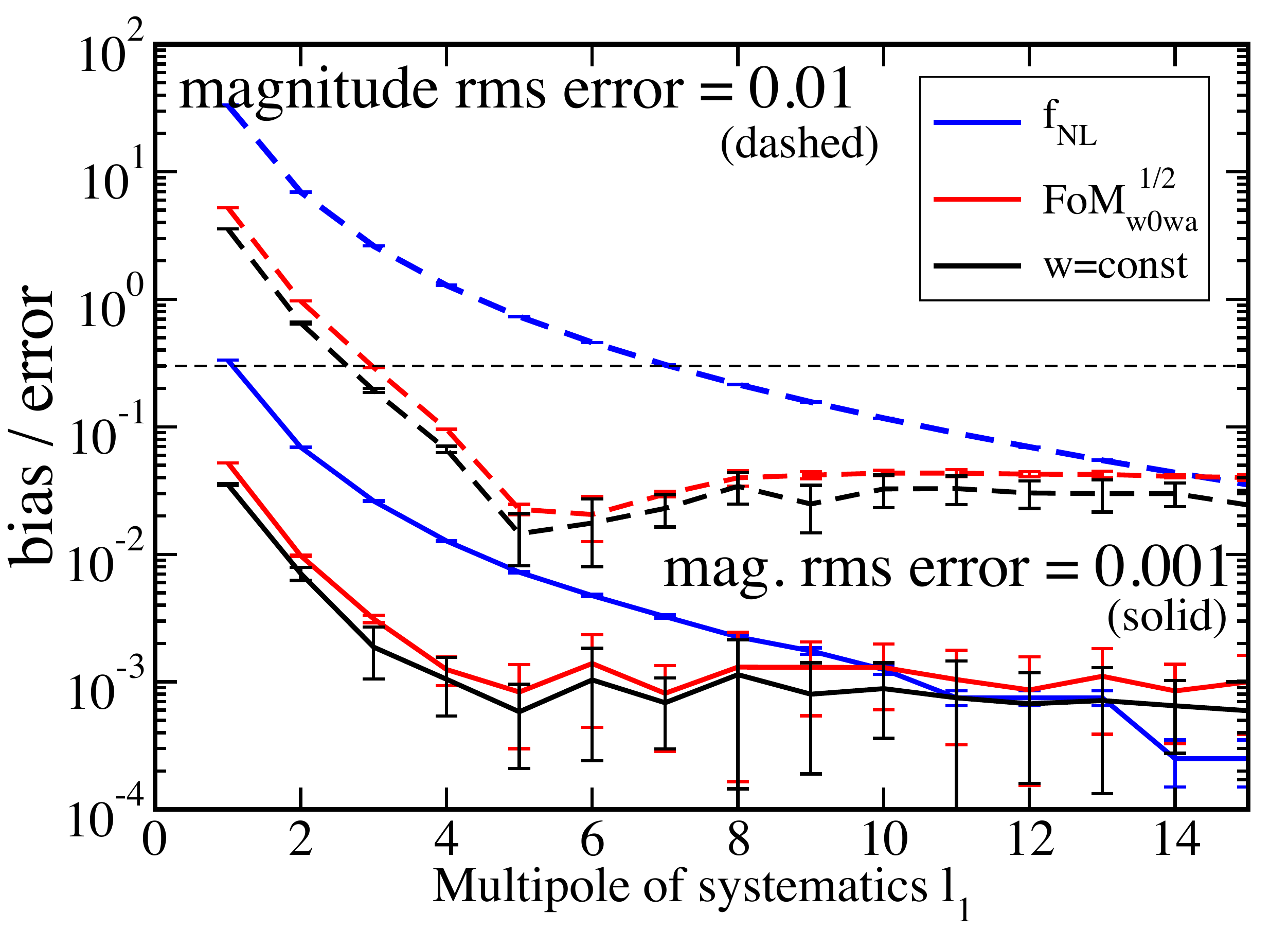} 
\caption{Bias divided by marginalized statistical error in the cosmological
  parameters for the fixed magnitude root-mean-squared variation of 0.001
  (solid curves) or 0.01 (dashed curves), as a function of multipole at which
  the systematics are introduced. We show the bias/error ratio for the
  non-Gaussianity parameter $\fnl$, constant equation of state of dark energy
  $w$, and the square root of the DETF figure of merit, $\FoM^{1/2}$, which
  serves to gauge any additional dependence brought forth by the temporal
  variation in the equation of state $w_a$.  To convert the magnitude
  variation to the $\delta N/N$ error, we used
  Eq.~(\ref{eq:dN_over_N_vs_magerr}) and the best-fit faint-end slope of the
  luminosity function, $s(z)$, estimated from simulations of
  \cite{jouvel2009}; see text for details.  The error bars show dependence on
  which of the $m_1$-values, for a fixed $\ell_1$, contains the calibration
  error; this dependence is small at the largest scales where the calibration
  error clearly has the largest effect. As discussed in the text, the monopole
  $\ell_1=0$ has no effect on the biases by definition. The dashed horizontal
  line denotes a fixed bias/error ratio of 0.3, which is approximately the
  upper limit of how much effect a systematic error should have on the
  cosmological parameters without seriously affecting the overall constraints
  in a survey.  }
\label{fig:biases}
\end{figure*}

The one important thing to take away from
Fig.~\ref{fig:Tl_minus_Cl_over_sigma} is therefore that the calibration error
is expected to be most damaging to the deepest surveys (or highest-redshift
slices of a survey). And it is precisely those highest redshifts that are most
valuable in providing information about dark energy and primordial
non-Gaussianity.

\begin{figure*}[t]
\centering
\includegraphics[scale=0.5]{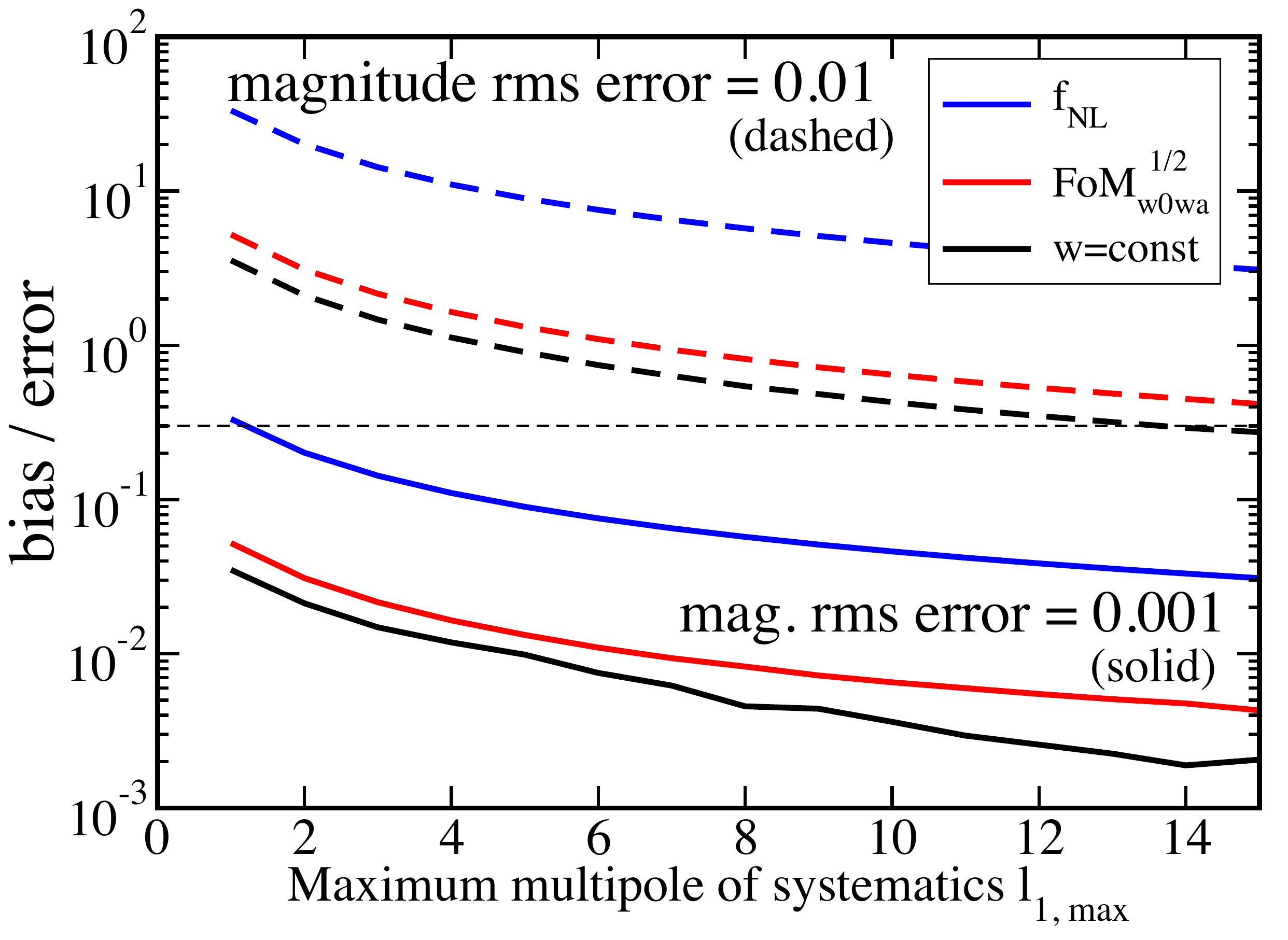} 
\caption{Same as Fig.~\ref{fig:biases}, except now the fixed magnitude error
  of 0.001 is {\it shared} equally among all multipoles in the range $1\leq
  \ell_1\leq \ell_{1, \rm max}$.  This is perhaps a more realistic assumption
  than the one shown in Fig.~\ref{fig:biases} where all of the error comes
  from a single multipole. The biases now appear larger because the
  contributions from the largest scales dominate the budget at a fixed
  $\ell_{1, \rm max}$.}
\label{fig:biases_mult_range}
\end{figure*}

\section{Sensitivity to calibration errors}\label{sec:sens}

Let us now consider a few specific examples. First, we will study the
sensitivities to an arbitrary systematic bias in calibration at each multipole
{\it separately}, i.e.\ one $(\ell_1, m_1)$ pair at a time in the $c_{\ell_1
  m_1}$. Then we study two concrete examples of physical effects that cause
calibration biases: corrections to the dust extinction maps, and variable
survey depth.

\subsection{From calibration to galaxy counts}

Consider the observed angular density of galaxies in some direction in the sky
$N(\nhat)\equiv \int n(z, \nhat)dz$, where $n(z,\nhat)$ is the galaxy density
in that direction and at redshift $z$.  Calibration errors correspond to
variations in the magnitude limit of the survey $\delta m_{\rm
  max}(\nhat)$. The observed density of galaxies changes since the galaxy
density is a strong function of the survey depth (i.e.\ the magnitude limit).
Suppressing the direction label $\nhat$, we have
\begin{eqnarray}
\delta [\log_{10} N(z, >m)] &=& \left .
{d\log_{10} N(z, >m)\over dm}\right |_{m_{\rm max}} \delta m_{\rm max}\nonumber\\[0.2cm]
&\equiv & s(z)\delta m_{\rm max}
\label{eq:sz}
\end{eqnarray}
where $m_{\rm max}$ is the maximal apparent magnitude observed in that
direction in some waveband. It follows that the systematic bias in the
observed fluctuations is
\begin{equation}
\left ({\delta N\over N}\right )_{\rm sys} = \ln (10) s(z)\, \delta m_{\rm max}.
\label{eq:dN_over_N_vs_magerr}
\end{equation}

Often we have information about the selective extinction $E_{B-V}$; the
relation to magnitude extinction is 
$\delta m\equiv \delta A = \delta[R\,E_{B-V}]$ where $R$ is the ratio of total
to selective extinction and $A$ is the alternative notation sometimes
  used for extinction. Assuming that $R$ is known perfectly\footnote{If $R$
  is not perfectly known, as is often the case, then the magnitude variation
  is equal to the variation in the product between $R$ and $E_{B-V}$.},
$\delta m \simeq R\, \delta (E_{B-V})$, and thus (restoring the direction
$\hat{\bf n}$ explicitly)
\begin{eqnarray}
\label{eq:dN_over_N_vs_sz}
\left ({\delta N\over N}\right )_{\rm sys}\!\!\!\!\!(\hat{\bf n}) \equiv 
c(\hat{\bf n}) 
&=& \ln (10) s(z) \delta m_{\rm max}(\hat{\bf n})\\[0.2cm]
&=& \ln (10) s(z) R\, \delta (E_{B-V})(\hat{\bf n}).
\label{eq:dN_over_N_vs_sz_with_R}
\end{eqnarray}
While $s(z)$ is galaxy-population dependent, we can still estimate $(\delta
N/N)_{\rm sys}$ to be {\it very} roughly of order $ \delta (E_{B-V})(\hat{\bf
  n})$, given that $s(z)$ is of order $0.1$-$1$ while $R$ 
takes values between about 1 and 5 depending on the band; see
e.g.\ the appendix of \citet{SFD}  and important updates given in the
Table 6 of \citet{Schlafly}.

\subsection{Calibration Bias per Multipole}\label{subsec:bias_per_mult}

Let us consider biases in our six cosmological parameters as a function of
bias in a single multipole $c_{\ell_1 m_1}$. Following the prescription in the
previous section, we assume that the variance of the calibration field is
fixed and constant separately at each multipole $\ell_1$; see Eq.~(\ref{eq:clm_var_const}).

We now propagate the calibration variation in a given $(\ell_1, m_1)$,
separately for $1\leq \ell_1\leq 20$ and $-\ell_1\leq m_1 \leq \ell_1$, and
for magnitude given in the above equations, to the observed angular power
spectra via Eq.~(\ref{eq:Tl}).

Figure~\ref{fig:biases} shows the bias divided by the statistical error in
cosmological parameters for the fixed {\it magnitude} rms variation per
multipole of $\langle \delta m_{\rm max}^2\rangle_{\rm sky}^{1/2}=0.01$ or
0.001. We use Eq.~(\ref{eq:dN_over_N_vs_sz}) to translate this to the
calibration-field variation, and then a modified version of
Eq.~(\ref{eq:clm_var_const}) to calculate the harmonic coefficients
$c_{\ell_1m_1}$ that enter the calculation:
\begin{equation}
c_{\ell_1 m_1}^{\rm Re, Im} = 
\left \{  \begin{array}{cl}
\sqrt{4\pi\Var(c(\hat{\bf n}))/(2\ell_1+1)}  \quad (m=0) \\[0.4cm]
\sqrt{2\pi\Var(c(\hat{\bf n}))/(2\ell_1+1)}  \quad (m\neq 0) \\[0.2cm]
\end{array} \right .
\label{eq:clm_var_const_all_m1}
\end{equation}
where, relative to Eq.~(\ref{eq:clm_var_const}), we have an additional term of
$(2\ell_1+1)^{-1/2}$ to keep the variance in each $\ell_1$ fixed since we are now
distributing power over {\it all} $m_1$-modes.  We adopt the fiducial
redshift-dependent faint-end slope of the luminosity function, $s(z)\equiv
\left .d\log_{10} N(z, >m)/dm\right |_{m_{\rm max}}$, of
\begin{equation}
s(z) =  0.094 + 0.155 z + 0.165 z^2
\label{eq:szfit}
\end{equation} 
estimated from the simulations of \cite{jouvel2009}, assuming a DES i-band
magnitude limit of 24.  This functional form roughly describes the trend that
the highest redshifts are most affected by variations in the survey depth.  We
emphasize that this form for $s(z)$ is meant purely for illustration, as
different galaxy samples will have different $s(z)$.  We consider biases in
the non-Gaussianity parameter $\fnl$, the (constant) equation of state of dark
energy $w$, and the square root of the dark energy figure-of-merit, which is
the inverse area of the 95\% contour in the $w_0$-$w_a$ plane
\cite{Hut_Tur_00,DETF}. Note that the latter quantity takes into account the
temporal variation of DE, and the square root serves to compare it fairly to
the bias in constant $w$; the two quantities, $\sigma(w)$ and $\FoM^{1/2}$,
show very similar behavior in these results.  The error bar at each $\ell_1$
shows the rms dispersion of the $(2\ell_1+1)$ values of $m_1$ into which we put
the systematics. So, for example, at $\ell_1=6$ and for either one of the rms
values for the calibration error, the error bars show the dispersion in the
bias/error ratios for $13$ different values of $m_1$.

Figure~\ref{fig:biases} clearly indicates that systematic errors have the
largest impact at largest angular scales assuming a fixed contribution to the
variance from each multipole. Bias in the non-Gaussianity parameter $\fnl$ is
larger than that for the dark energy parameters, which is expected because
most of the information on $\fnl$ comes from large angular scales which are
particularly susceptible to calibration errors. For the calibration
systematics at smaller scales -- $\ell_1$ beyond six or so corresponding to
variation at scales less than about 30 degrees on the sky -- the effect of the
systematics asymptotes to a smaller value. The minimum in the $\FoM^{1/2}$ and
$w$ curves around $\ell_1\simeq 6$ is due to the transition from the dominance
of the additive errors at larger scales to multiplicative errors at smaller
scales. Since calibration errors at large angular scales are the most
damaging, it is sufficient to consider only those, and our choice of the
maximum multipole of where the error enters, $\ell_1\leq \ellcalibmax = 20$ is
therefore sufficient.

Figure \ref{fig:biases_mult_range} is similar to Fig.~\ref{fig:biases} except
now we show the effects when the magnitude error is split equally among all
multipoles less or equal to $\ell_{1, \rm max}$ (instead of all of it being
lumped in a single multipole as in Fig.~\ref{fig:biases}).  Given that the
effect of the systematic error decreases with $\ell_1$, the biases in the
cosmological parameters are larger than in the previous figure.
Qualitatively, the two figures paint a consistent picture of the potentially
deleterious effects of the calibration variations even at a level
corresponding to $O(0.001$--$0.01)$ magnitudes.

\subsection{Example I: Corrections to dust maps}\label{sec:PG10}

We now study a specific scenario of calibration systematics: corrections to
the SFD \cite{SFD} dust extinction maps.  Dust in our Galaxy causes
extinction, which in turn alters the observed galaxy fluctuations across the
sky. While the Galactic dust has been mapped out reasonably accurately, we ask
how accurately it {\it needs to} be mapped out in order not to bias the
cosmological parameters.

\begin{figure*}[t]
\centering
\includegraphics[scale=0.32]{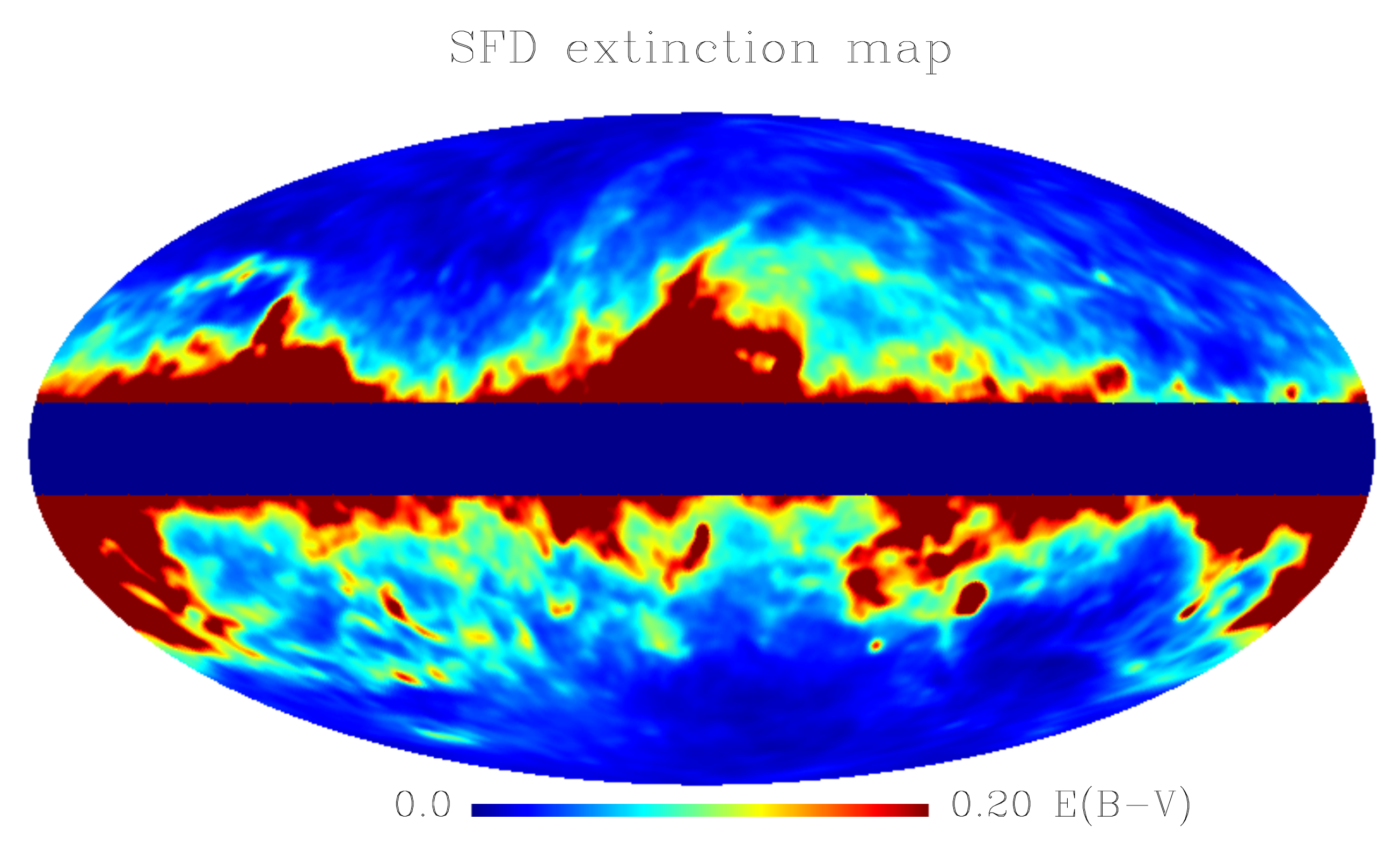} 
\includegraphics[scale=0.32]{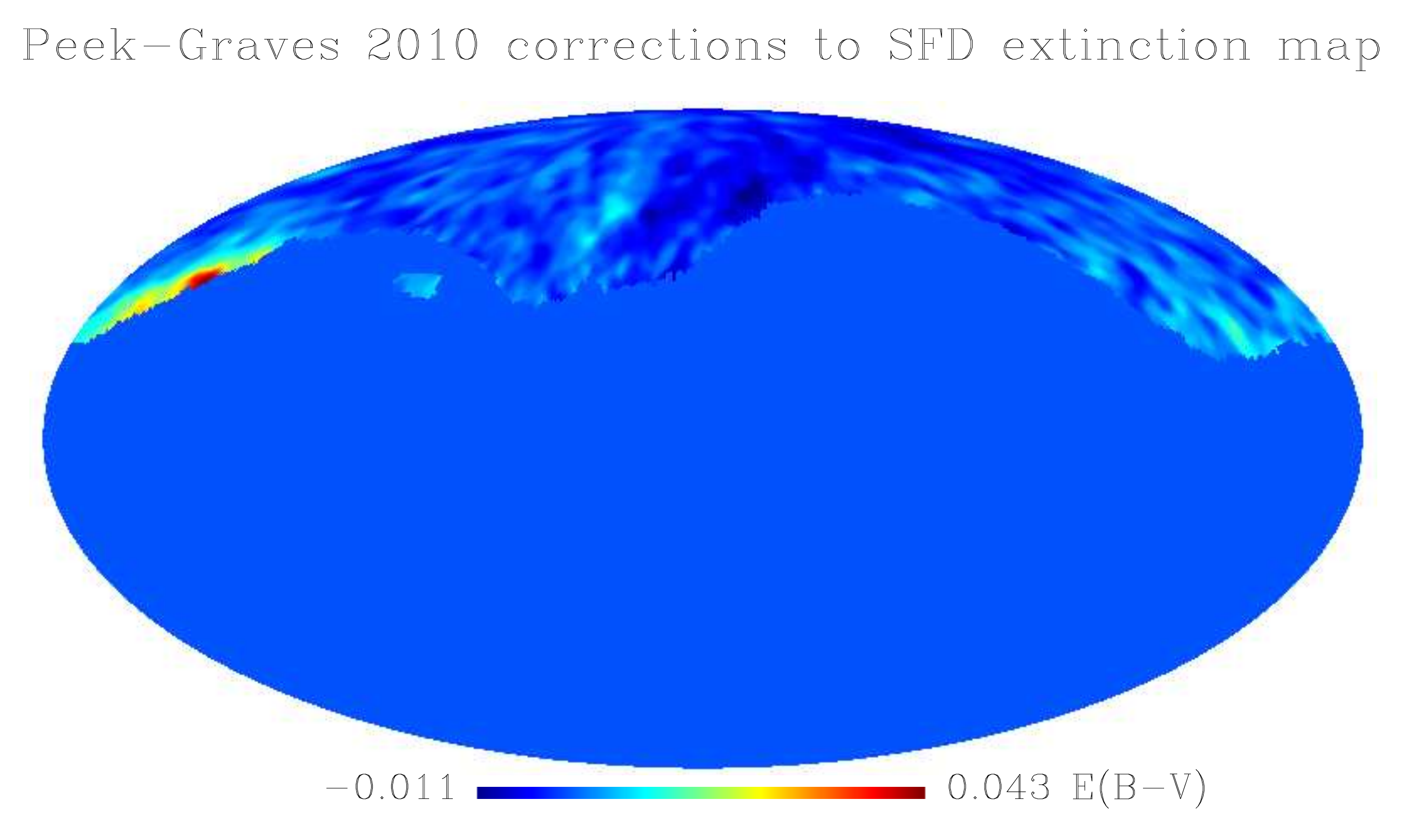} 
\includegraphics[width=0.45\textwidth]{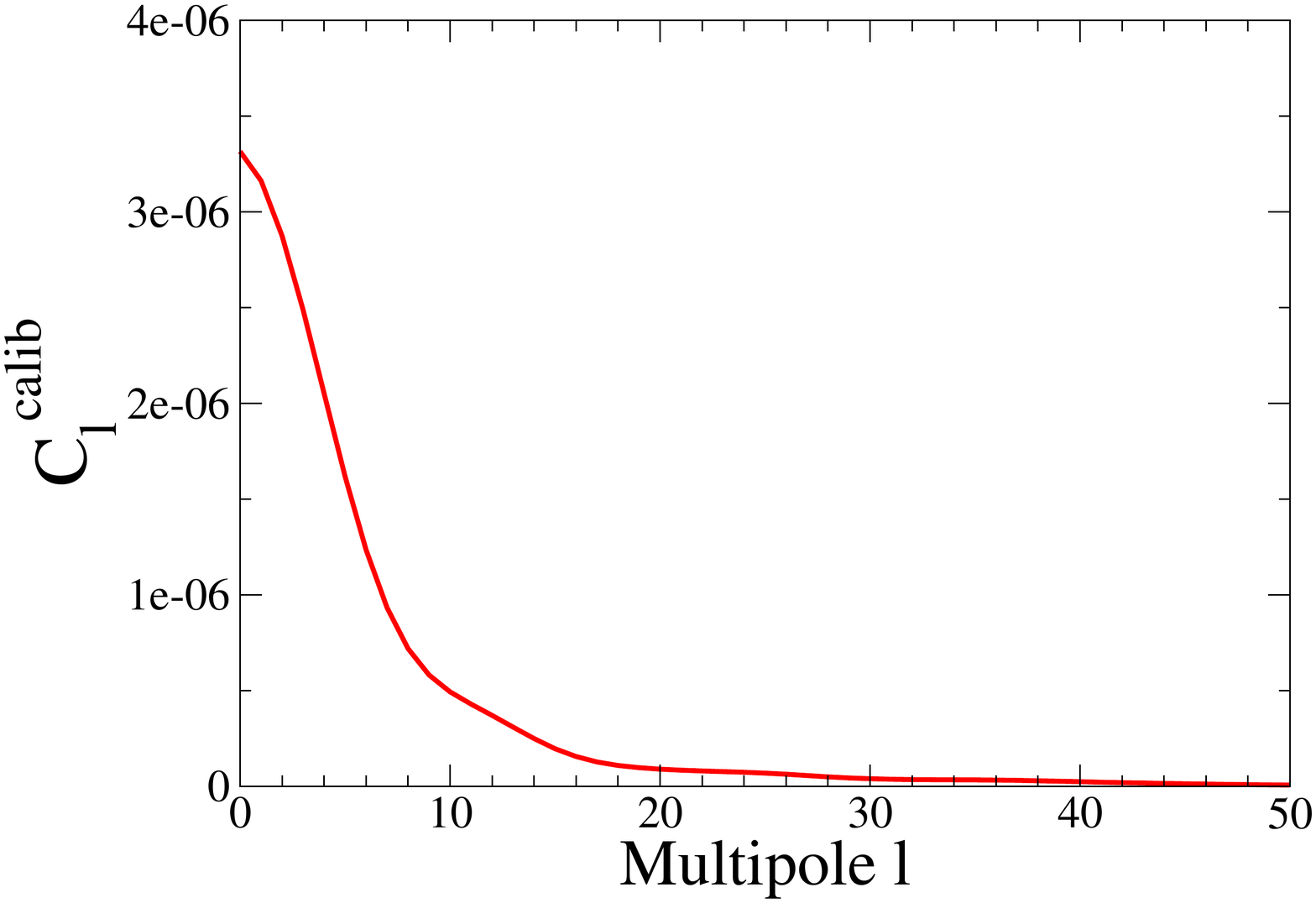} 
\includegraphics[width=0.45\textwidth]{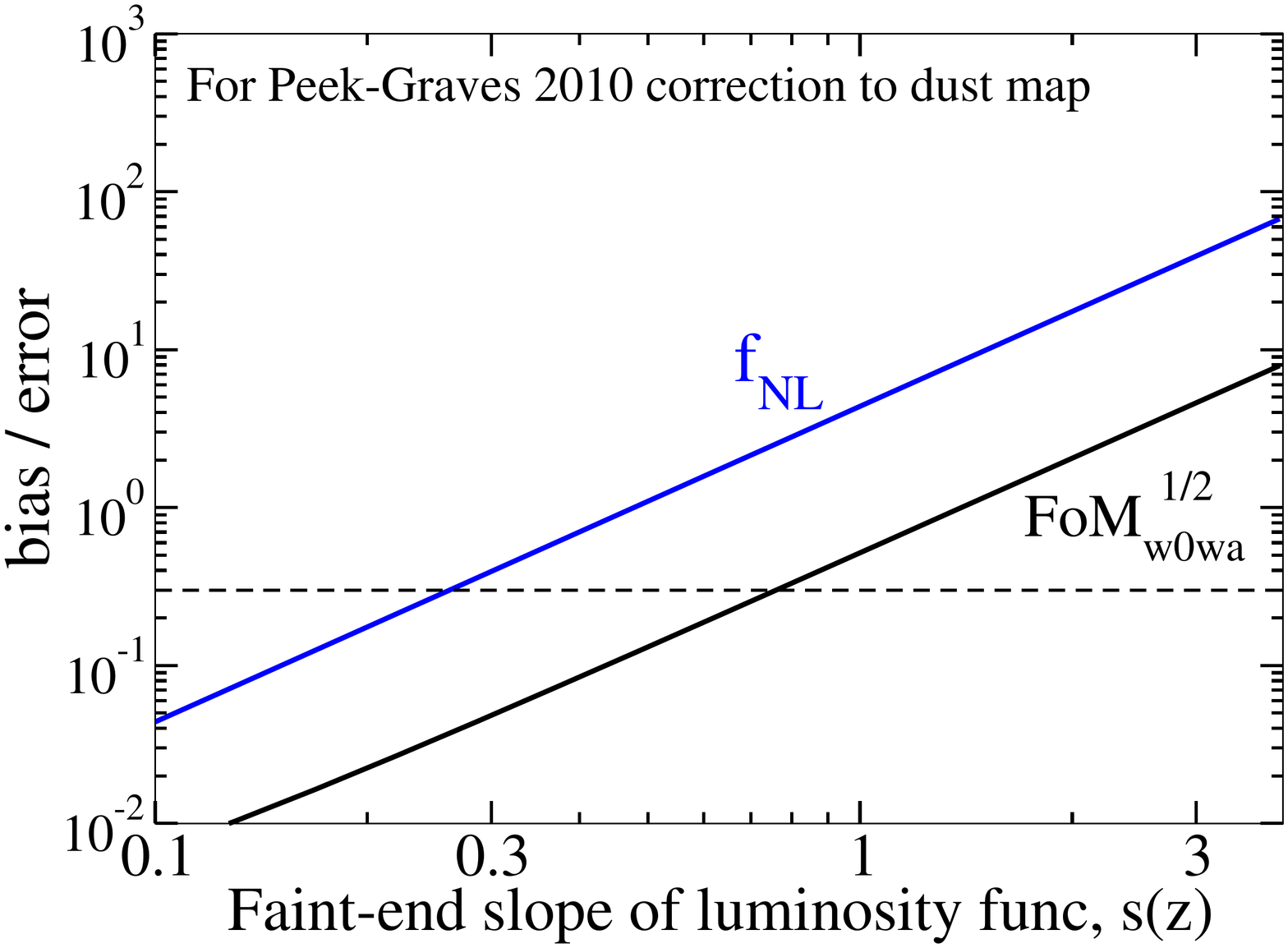} 
\caption{Top left: \citet{SFD} SFD extinction map, $E_{B-V}(\nhat)$, in
  Galactic coordinates with the 10 degrees Galactic plane cut. Top right:
  corrections to the SFD map from the work of \citet{Peek_Graves}. Bottom
  left: angular power spectrum of the PG10 map extracted by {\tt Polspice} and
  shown without the usual $\ell(\ell+1)/(2\pi)$ term so that the relative
  contribution of different multipoles can be more easily seen. Bottom right:
  bias/error ratios for $\fnl$ and the square root of the DETF FoM assuming
  PG10 map represents the calibration error, as a function of the faint-end
  slope of the luminosity function $s$. Note that the biases increase very
  sharply with $s$, roughly scaling as $s^2$. The desired bias/error limit
  (horizontal dashed line) is exceeded already for $s\simeq 0.3$ for $\fnl$
  and $s\simeq 0.8$ for the dark energy equation of state.}
\label{fig:PG10}
\end{figure*}

\begin{figure*}[th]
\centering
\includegraphics[width=0.45\textwidth]{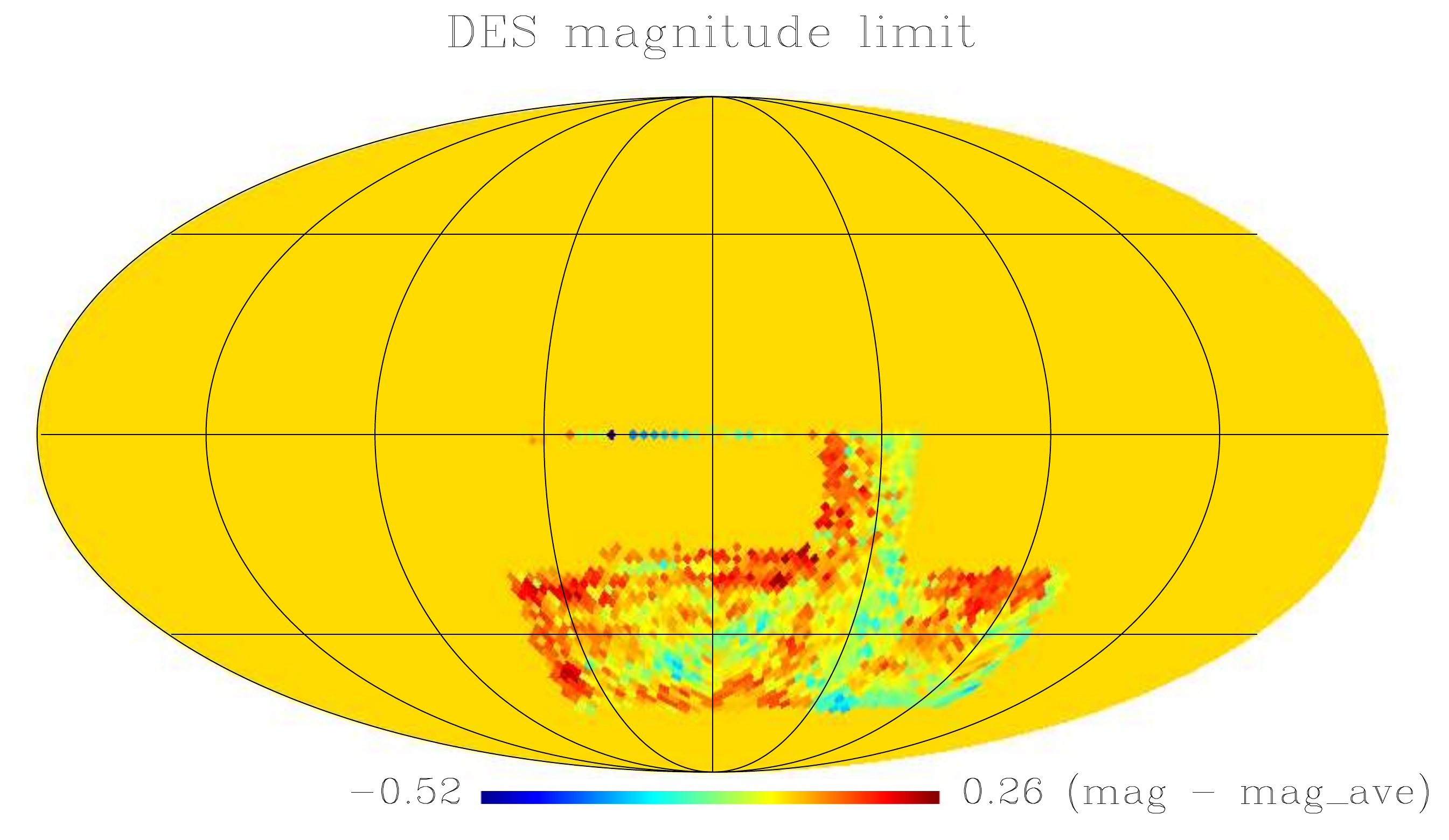} \\
\includegraphics[width=0.45\textwidth]{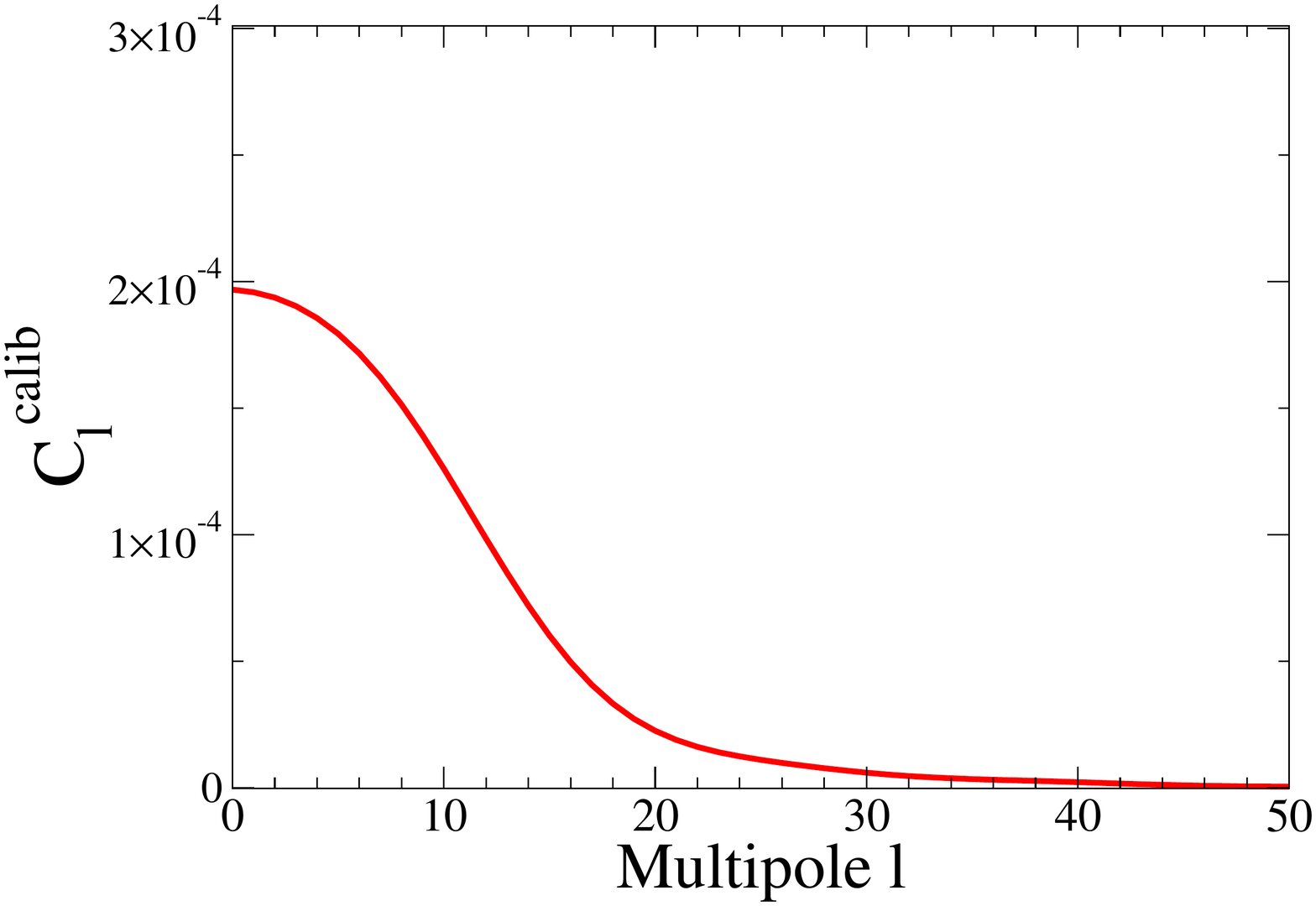} \hspace{0.1cm}
\includegraphics[width=0.45\textwidth]{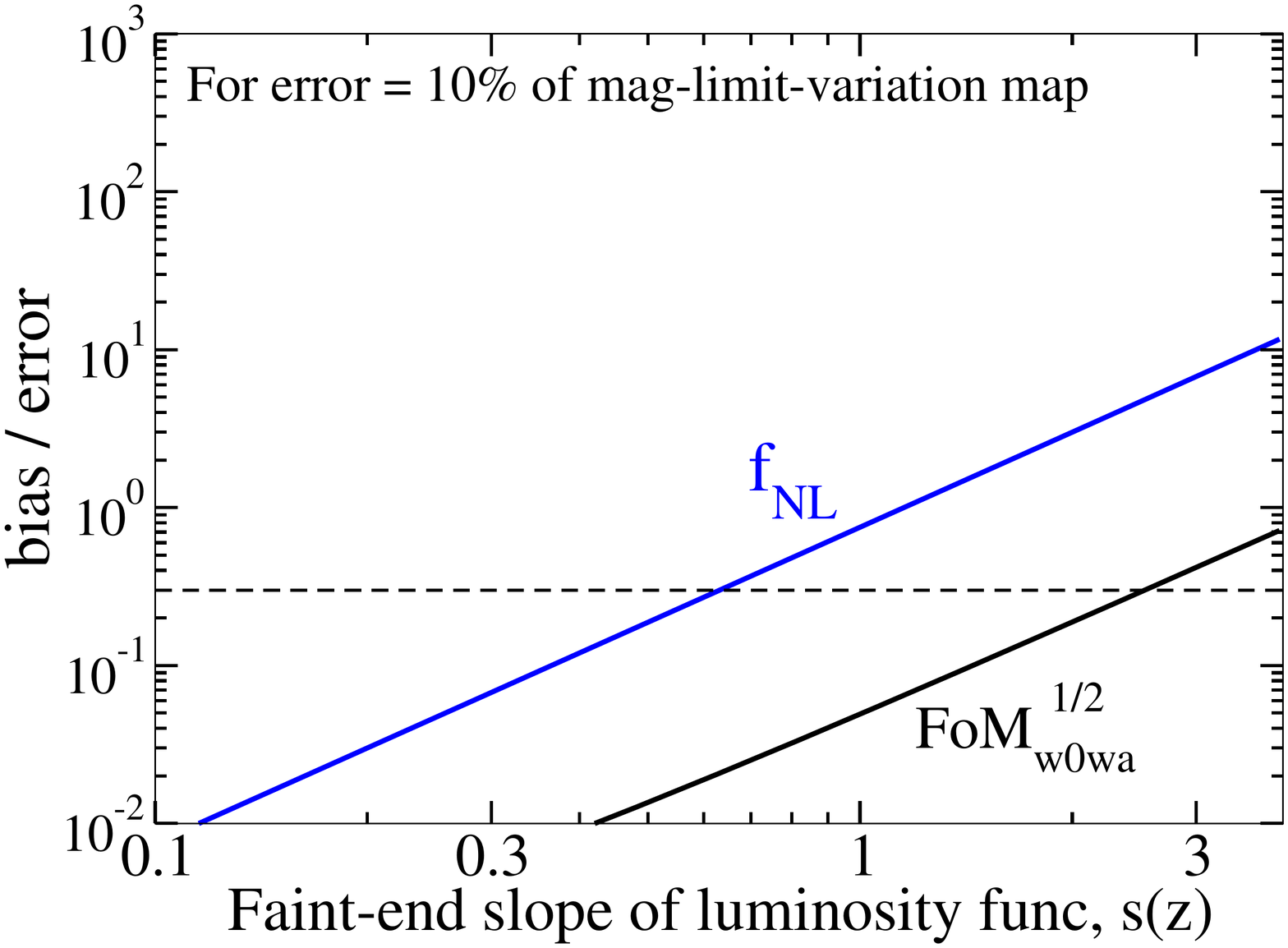} 
\caption{Top panel: i-band magnitude limits estimated for the upcoming
  observations of the Dark Energy Camera at CTIO as a function of angular
  position.  The pattern of variations in the magnitude limits are set by the
  variations in the observing conditions and the survey tiling strategy over
  the five years of the survey.  Bottom left: angular power spectrum of the
  magnitude limi map, extracted using {\tt Polspice} and shown without the
  usual $\ell(\ell+1)/(2\pi)$ term so that the relative contribution of
  different multipoles can be more easily seen. Bottom right: biases in the
  cosmological parameters vs.\ the faint-end slope of the luminosity function
  $s(z)$ assuming calibration error maps is consistent with a fixed {\it
    fraction} of 10\% of amplitude (or 1\% of power) of the magnitude-limit
  map shown in the top (bottom left) panel. The desired bias/error limit
  (horizontal dashed line) is exceeded for $s(z)\simeq 1$. }
\label{fig:Annis}
\end{figure*}

To start, we need a model for the variations in the SFD map. We adopt results from the work of
\citet{Peek_Graves} (hereafter PG10) who used 'standard crayons' -- objects of known color --
in the SDSS to correct the SFD maps over the north galactic cap region (for a
related work, see \citet{Schlafly}). Schematically, therefore
\begin{equation}
\mbox{calibration variations $\equiv$ (PG10 -- SFD)}.
\nonumber
\end{equation}
The SFD map is shown in the top left panel of Fig~\ref{fig:PG10}, while the
PG10 {\it correction} is displayed in the top right panel.  To convert this
$E_{B-V}$ map to $\delta N/N$ fluctuations (see
Eq.~(\ref{eq:dN_over_N_vs_sz_with_R})), we assume DES observations in the
i-band, for which $R=1.595$ \cite{Schlafly}.

The bottom left panel in Fig.~\ref{fig:PG10} shows the angular power spectrum
extracted using {\tt Polspice} software package \cite{polspice}. As explained in Appendix
\ref{app:reconstr} at length, the most reliable way of modeling the
calibration errors was to first extract the power from the map, then generate
a full-sky realization {\it consistent} with that power (using the {\tt
  isynfast} routine in HEALPix). We explicitly verified the intuitive
expectation that the results do not depend much on the realization. 

The bottom right of Fig.~\ref{fig:PG10} shows the resulting biases in $\fnl$
and the square root of the dark energy FoM, as a function of the faint-end
slope of the luminosity function $s(z)$; the $w={\rm const}$ case is not shown
here or in the following Figure since it gives very similar results as the
$\FoM^{1/2}$. We assume calibration variations are given by the PG10
corrections, and that $s(z)$ is constant in redshift (we nevertheless allow
for the redshift-dependent $s(z)$ in all equations). As mentioned around
Eq.~(\ref{eq:dN_over_N_vs_magerr}), the biases are very sensitive to
$s(z)$, scaling very nearly as $s(z)^2$. This can be easily understood: in
Sec.~\ref{sec:bias_isotropic} we mentioned that the bias in the power spectrum
is dominated by the added calibration power $\propto |\clm|^2$, while the
(real-space) calibration field is linear in $s(z)$
(Eq.~\ref{eq:dN_over_N_vs_sz}); hence
\begin{equation}
\delta p_a\propto \delta C_\ell\propto |\clm|^2\propto s(z)^2.
\end{equation}
In other words, in the case where the additive calibration errors dominate so
that calibration simply adds power ($|\clm|^2$ term), the biases in the
cosmological parameters are proportional to this added power, and hence to the
square of the faint-end slope of the luminosity function.  Therefore, the
faint-end slope of the luminosity function is a key factor relating the
photometric magnitude variations to the cosmological parameter biases. Steep
faint-end slopes will lead to particularly stringent requirements on our
understanding of the large-angle photometric variations in the survey.
Regardless of the value of the faint-end slope, however, the bottom right
panel of Fig.~\ref{fig:PG10} shows that the effects of the imperfectly
estimated Galactic dust on the cosmological parameters can be very significant.

\subsection{Example II: variability of survey depth}\label{sec:Annis}

Our second example is based on the expectations of photometric depth
variations of the Dark Energy survey.  We use a map (Jim Annis, private
communication) simulating observations over 525 night of observation spread
over five years; see the top panel of Fig.~\ref{fig:Annis}.  The observing
conditions on the site are based on historical atmospheric data of the CTIO
site between 2005 and 2010.  The tiling strategy uses multiple massive
overlaps to generate a survey that is as homogeneous as possible.  Each part
of the sky is imaged ten times in each of the five DES filters (grizY).  For
simplicity, we only focus on the i-band survey-depth map.

The effect of the unaccounted-for variability in the survey depth is the same
as that of the photometric calibration error.  However, the variability is
large and expected to be taken into account; therefore, we (arbitrarily) adopt
the final calibration error to be equal to one-tenth of the depth-variation
map (i.e.\ 1/10 of its amplitude shown in the top panel of
Fig.~\ref{fig:Annis}). In other words, we assume
\begin{equation}
\mbox{calibration variations $\equiv  {1\over 10}\times$(i-band variability
  map)}.
\nonumber
\end{equation}

We follow the same procedure as with the dust example above, and calculate the
power spectrum of the depth variability map using  {\tt Polspice};
see the bottom left panel of Fig.~\ref{fig:Annis}. 
The variability of the
survey depth will of course be accounted for in the data analysis -- if it
were not, it would lead to large biases in cosmological parameter estimates (as
we easily verified using our formalism). The question is, then, to what
accuracy do these variations need to be understood?  

We answer that question by plotting, in the bottom right panel of
Fig.~\ref{fig:Annis}, the bias in the (square root of the) DE FoM, and
non-Gaussianity parameter $\fnl$, as a function of the faint-end slope of the luminosity
function $s(z)$. As in the previous example of the corrections to the SFD dust
maps, we find that the biases in the cosmological parameters are significant,
and that they strongly depend on the faint-end slope of the luminosity
function. In fact, even assuming that only 10\% of the variability in the
survey depth is the ``calibration error'' -- the case shown in the bottom
right panel of the Figure -- the bias/error ratios are still large if
$s(z)\gtrsim O(1)$.

\section{Conclusions}\label{sec:concl}

In this paper we made a first fully general study of the effect of the
photometric calibration variations on the measured galaxy clustering angular
power spectra. We derived a general formula for how a calibration variation
with arbitrary spatial dependence affects the measured galaxy angular power
spectrum.  We illustrated the results assuming the standard set of
cosmological parameters (including $\fnl$), DES-type dataset with five
tomographic bins out to $\zmax=1$, and two specific examples of real-world
photometric calibrations. We now summarize our findings.

Photometric variations modulate the observed angular distribution of galaxy
counts according to Eq.~(\ref{eq:nobs}).  This modulation translates into
additive and multiplicative changes to the observed density fluctuation field,
cf. Eqs.~(\ref{eq:deltaobs}) and (\ref{eq:tlm}), which in turn generate
additive and multiplicative changes to the observed power spectrum.

As shown in Eq.~(\ref{eq:tlm_tl'm'}), photometric variations across the survey
masquerade as apparent violations of statistical isotropy.  Hence, explicit
tests of statistical isotropy could provide a useful way to identify
unaccounted-for variations in the photometry.  In this paper, we focused on
the effects in the angle-averaged power-spectrum, cf.\ Eq.~(\ref{eq:Tl}).  We
found that large-angle modulations of power (dipole, quadrupole, etc), are
particularly damaging to cosmological analysis.  We demonstrate this
explicitly (cf.\ Eq.~(\ref{eq:tl-cl}) and Fig.~\ref{fig:Tl_minus_Cl_over_sigma})
for the case where the variance in the photometric calibration error field is
concentrated in one multipole $\ell_1$ at a time.  Note that the spatially
uniform photometric decrement or increment across the sky (i.e.\ the monopole,
$\ell_1=0$) is unobservable since it only affects the mean number of galaxies
in the survey.

Specializing in the angle-averaged power spectrum as done in
Eq.~(\ref{eq:Tl}), one can explicitly show that largest-angle fluctuations are
dominant (for a fixed induced variance on the calibration error field
$c(\hat{\bf n})$); see Fig.~\ref{fig:Tl_minus_Cl_over_sigma}. Moreover,
highest-redshift clustering measurements are most susceptible to the
photometric variations, essentially because their angular power is the
smallest and is therefore most affected by the photometric variation.

Less obviously, we find that the additive errors (e.g.\ term proportional to
$|\clm|^2$ in Eq.~(\ref{eq:Tl})) are typically dominant over the
multiplicative biases (terms proportional to the coefficients $U$) for all
redshift bins and at large angular scales. The reason is simple: because they
couple different multipoles, multiplicative terms are suppressed relative to
the additive ones by the fiducial angular power spectrum $\Cl$ factor; see the
term with $C_{\ell_2}$ in Eq.~(\ref{eq:Tl}). Since $\Cl\ll 1$ even at low-z
(and all $\ell$), the additive terms dominate the error budget if all $\ell$
modes are used in the analysis.  However, at slightly smaller angular scales
($\ell\gtrsim 10$) the multiplicative error terms dominate the error budget
and can significantly bias the cosmological constraints, as discussed in
Sec.~\ref{sec:sens}.  Therefore, it is important to include both
multiplicative and additive aspects of the calibration error to accurately
model biases in cosmological analyses.

The photometric variation calibration errors affect the galaxy clustering
signal at large spatial scales, and lead to biases in the inferred
cosmological parameters. Parameters describing dark energy and, especially,
primordial non-Gaussianity are particularly affected since they imprint
signatures in the clustering of galaxies precisely at these large scales.
Figures \ref{fig:biases} and \ref{fig:biases_mult_range}, the principal plots
in this paper, show these cosmological parameter biases for a fixed
contribution of the calibration error at each multipole separately and for a
range of multipoles, respectively. In Sec.~\ref{sec:sens} we further give two
specific real-world examples of what the photometric variations could occur:
errors in mapping the dust in our Galaxy, and variations in survey depth. We
find that these calibration errors lead to potentially large cosmological
biases, especially if the faint-end slope of the luminosity function $s=d\log
N/dm|_{m_{\rm max}}$ is steep.  In particular, in the Fisher matrix
approximation, the cosmological parameter biases scale as $s^2$.

As a by-product of this work, we developed a reasonably fast algorithm, and
provide a
code\footnote{\url{www-personal.umich.edu/~huterer/CALIB_CODE/calib.html}}, to
calculate the full biased angular power spectrum of galaxies given an
arbitrary photometric calibration variation map on the sky. As mentioned
above, this calculation {\it would} be rather trivial if we could assume that
the power spectrum errors are purely additive; however we demonstrated that
multiplicative errors are important and may in fact dominate, necessitating
the more numerically intensive calculation for surveys of interest.

From our analyses, it appears that the total rms of the calibration error has
to be kept at the level at somewhere between 0.001 and 0.01 magnitudes,
depending on how large a scale one wants to consider in order to maximize
extraction of cosmological information, in order not to bias the cosmological
parameters appreciably.  This is a very stringent requirement!  Achieving it,
however, may not be as difficult as it sounds given that this is the
time-averaged error to be tolerated at the end of the whole survey. Moreover,
there are several other tools that we have at our disposal that we did not
consider in this preliminary work. For example, one could use the survey
itself to internally determine (``self-calibrate'') the photometric-variation
errors, similarly to what weak lensing, cluster count, or type Ia supernova
surveys are doing or planning to do with their systematic errors.  One could
also use measurements of the higher-point correlation functions to help
determine these nuisance parameters (in our language, the $c_{\ell_1
  m_1}$). We leave these promising avenues for future study.

\section*{Acknowledgements}
We thank Jim Annis for providing the expectations of photometric depth
variations for the DES, and Joshua Frieman, Enrique Gaztanaga, Andrew Hearin,
Shirley Ho, Will Percival, Ashley Ross, Eddie Schlafly and Daniel Shafer for
useful discussions. We also thank the anonymous referee for a number of useful
suggestions. CC and DH are supported by the DOE OJI grant under contract
DE-FG02-95ER40899. CC is also supported by a Kavli Fellowship at Stanford
University. DH is additionally supported by NSF under contract AST-0807564,
and NASA under contract NNX09AC89G. WF is supported by NASA under contract
NNX12AC99G.

\appendix

\section{Angular power spectra and their information content}\label{app:PS}

We model model the angular power spectra of galaxy density fluctuations as
follows. In Limber's approximation, the angular power spectrum is given by
\begin{equation}
P_{\ell}^{(ij)}  =  {2\pi^2\over \ell^3}\int_0^\infty dz \, r(z)  H(z) W_i(z) W_j(z)  
\Delta^2\left (k={\ell\over r(z)}, z\right ),
\end{equation}
where $\Delta^2(k)\equiv k^3P(k)/(2\pi^2)$ is the dimensionless power
spectrum, $r(z)$ is the comoving angular diameter distance and $H(z)$ is the
Hubble parameter. The weights $W_i$ are given by
\begin{equation}
W_i(z)=\frac{n(z) }
{\int_{\zmin^{(i)}}^{\zmax^{(i)}}n(z) dz}\, [H(\zmax^{(i)})-H(\zmin^{(i)}) ],
\end{equation} 
were $H(x)$ is the Heaviside step function and $\zmin^{(i)}$ and $\zmax^{(i)}$
are the lower and upper bound of the $i$th redshift bin and $n(z)$ is the
normalized radial distribution of galaxies\footnote{Note that a sometimes-used
  alternative definition of $n(z)$ refers to the spatial density of galaxies
  (e.g.~\citet{hu2004joint}); it is related to the quantity we use via $dN/dz
  = n(z)\,\Omega\,r^2(z)/H(z)$, where $\Omega$ is the solid angle spanned by
  the survey, and $r$ and $H$ are the comoving distance and Hubble parameter,
  respectively. Note also that our $W(z)$ is equivalent to the quantity $f(z)$
  from \citet{ho2008correlation}.}  whose form we take to be
\begin{equation}
n(z) = {1\over 2z_0^3}\, z^2e^{-z/z_0},
\end{equation}
which peaks at $\zmax=2z_0$. In this work we use $z_0=0.3$ so that the radial
distribution peaks at redshift 0.6.

At the lowest multipoles the Limber approximation is not accurate any more;
see Fig.~\ref{fig:Cl}. At $\ell\leq 30$ we adopt the full expression for the
power spectrum; using notation from e.g.~\cite{Hearin_Gibelyou_Zentner} this
is
\begin{equation}
\begin{aligned}
P_{\ell}^{(ij)} &= 4\pi\int_0^\infty d\ln k\, \Delta^2(k, z=0)I_i(k)I_j(k)\\[0.2cm]
I_i(k) &\equiv  b_i(k) \int_0^\infty dz\, W_i(z) {D(z)\over D(0)}\, j_\ell(k\chi(z))
\label{eq:noLimber}
\end{aligned}
\end{equation}
where $b_i(k)$ is the bias in $i$th redshift bin, $\chi(r)$ is the radial
distance, and $\chi(z)=r(z)$ in a flat universe that we consider. Here $D(z)$
is the linear growth function of density fluctuations, so that
$\delta(z)=[D(z)/D(0)]\delta(0)$. At multipoles $\ell>30$ we continue to use
the Limber approximation since the computation is much faster. The power
spectra are shown in Fig.~\ref{fig:Cl}. Note that, when using
Eq.~(\ref{eq:noLimber}), the non-linear corrections to clustering are
negligible since we only apply this equation at very large scales.

\begin{figure}[h]
\centering
\includegraphics[width=\linewidth]{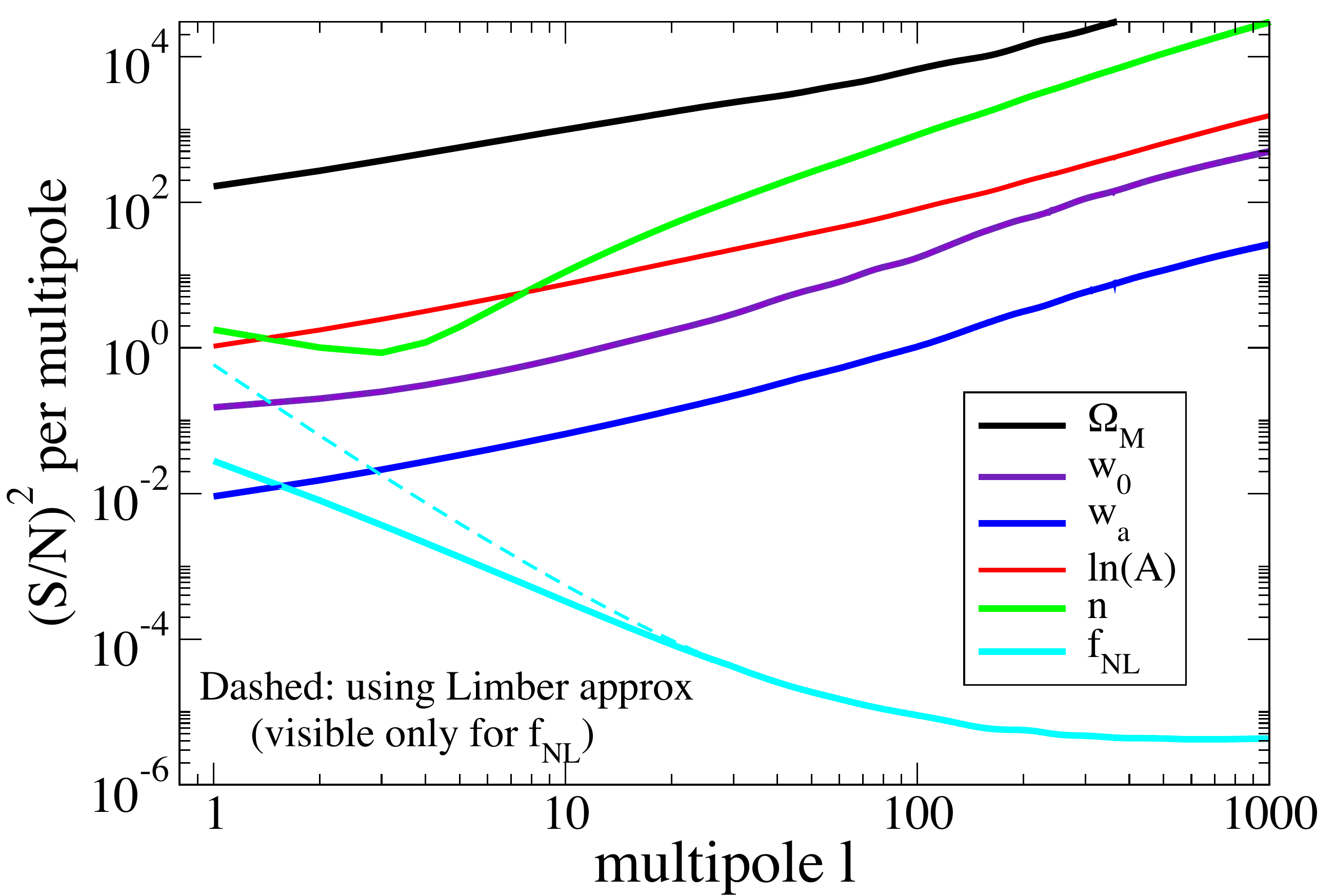} 
\caption{Signal-to-noise squared contribution (i.e.\ contribution to diagonal
  Fisher matrix element with no systematics present) per single multipole
  $\ell$ for each cosmological parameter. Note that we use the Limber
  approximation only at $\ell>30$ in order to more accurately capture
  information on $\fnl$ at $\ell\leq 30$.}
\label{fig:SN_per_mult}
\end{figure}

Figure \ref{fig:SN_per_mult} shows the contribution to the signal-to-noise squared
for each cosmological parameter separately per multipole -- in other words, we show
contribution to the Fisher matrix element $F_{aa}$ for each parameter $p_a$
from a single multipole $\ell$. There is a clear trend of increased
information at high multipoles, except for $\fnl$ whose information largely
comes at the lowest multipoles, despite cosmic variance \citep{Dalal}. 

\section{Extraction of full-sky calibration from cut-sky data}
\label{app:reconstr}

In order to estimate the effect of photometric calibration errors, we perform
all calculations in multipole space. In particular, our formalism requires the
multipole coefficients of the calibration variation field, $c_{\ell_1 m_1}$,
as input. In practice, this job will be left to observers who will have a
number of tools at their disposal to estimate the photometric variation, as a
function of sky position, in their survey. They, or their theory colleagues,
can then use our formalism, Eq.~(\ref{eq:Tl}), to estimate the effect of
calibration variations on the angular power spectra and the cosmological
parameters.

However it is nontrivial to obtain the coefficients $c_{\ell_1 m_1}$ given the cut-sky
observations, i.e.\ partial sky coverage: recall, for example, that the Peek-Graves
dust-correction map from Sec.~\ref{sec:PG10} covers about 1/4 of the sky, while
the DES depth map from Sec.~\ref{sec:Annis} covers only 1/8 of the sky. 
Note that the naive reconstruction from the observed map, $c_{\ell_1 m_1} =
\int Y^*_{\ell_1 m_1}({\hat{\bf n}}) \delta N/N ({\hat{\bf n}})d\Omega$, would
give the {\it cut-sky} multipoles which describe the field that is zero
outside of the observed region, which we do {\it not} want: the cut-sky
multipoles have wrong amplitudes and harmonic structure relative to the
full-sky 'truth'. 

Filling in the missing sky is a well-known problem in the cosmic microwave
background literature. While the reconstruction of the power $\Cl$ is
relatively straightforward, reconstruction of the temperature field $\alm$ is
challenging   and typically works well only when substantial portions of the sky
($\fsky\gtrsim 0.8$) are observed. Fortunately, in this work we are only
interested in obtaining {\it approximate} values for the $c_{\ell_1 m_1}$,
consistent with the cut-sky map of the calibration field, in order to estimate
its effects on cosmological parameters.

We have attempted, without success, two rather well-known techniques for
reconstructing the true, full-sky $c_{\ell_1 m_1}$ from partial sky observations:

\begin{itemize}
\item We tried the direct reconstruction of the field, using the
  maximum-likelihood procedure (see e.g.\ \cite{Efstathiou2004-hybrid,
    CMB-mapmaking,bias_Copi_reconstr}); this approach is used, for example, to
  reconstruct the CMB power at $\ell\lesssim 30$ \cite{wmap7}. While this
  approach returns the true structure in the observed area of the sky (as it
  must), the reconstructions add spurious structure and extra
  power -- clearly visible by eye -- in the unobserved areas of the sky. 
\item We tried ``harmonic inpainting'' \cite{Hoffman_Ribak, Kim_inpainting} of
  our two maps, where the missing portion of the sky is filled in assuming that
  the density field is Gaussian, and starting with a guess for the covariance
  matrix of the field (which is given in terms of the theoretical angular
  power spectrum). As with the direct reconstruction, we  find  large spurious
  power in the unobserved portion of the sky in realistic cases when sky
  coverage is small, $\fsky\ll 1$. 
\end{itemize}

Given these failures and the fact that we only need a rough estimate of the
full-sky calibration field $c_{\ell_1 m_1}$, we resort to the simpler scheme
of reconstructing the large-angle power spectrum and drawing realizations of
the field {\it consistent} with it. In particular, we
\begin{enumerate}
\item Calculate the angular power spectrum of the calibration field, $\Cl^{\rm
  calib}$, from the cut-sky calibration map using {\tt Polspice} package
  \cite{polspice}. {\tt Polspice} uses the pixel-based approach to calculate
  the real-space angular power first and then, with appropriate apodizations,
  convert it to the $\Cl^{\rm calib}$.
\item Sample from this angular power spectrum, i.e.\ $c_{\ell_1
  0}=\mathcal{N}(0, C_{\ell_1}^{\rm calib})$ and $c_{\ell_1 m_1}^{\rm Re,Im}=\mathcal{N}(0,
  C_{\ell_1}^{\rm calib}/2)$ for $m_1\neq 0$, where $\mathcal{N}(\mu, \sigma^2)$ is the
  Gaussian normal function with mean $\mu$ and variance $\sigma^2$.
\item Repeat the previous step a number of times to ensure appropriate
  averaging over realizations, though in practice we find that the each
  realization (for a fixed estimated $C_{\ell_1}^{\rm calib}$) leads to similar results.
\end{enumerate}
The approach returned a smooth power spectrum at large scales (see the $\Cl^{\rm calib}$
panels in Figs.~\ref{fig:PG10} and \ref{fig:Annis}), and therefore reliable
multipole coefficients describing calibration variations across the whole sky. 

The reader might be worried that, in the above-mentioned successful approach,
we assumed Gaussianity for the calibration field by assuming that the
$C_{\ell_1}^{\rm calib}$ describes the full-sky field, and moreover assuming that the
$c_{\ell_1 m_1}$ are Gaussian random variates. This is definitely an
assumption needed to get the coefficients {\it in our particular examples}
described in this Appendix but not in general. Moreover, we are not concerned
about the assumption of Gaussianity of the calibration-variation field:
Eq.~(\ref{eq:tlm_tl'm'}), for example, illustrates that the calibration field
will break statistical isotropy of the observed density fluctuations
regardless of the nature of the statistical distribution of the ensemble from
which th $c_{\ell_1 m_1}$ come from. Modest dispersion among the different
$m_1$-modes in Fig.~\ref{fig:biases} confirms that the principal property of
these coefficients that drives their effect on cosmology are their amplitudes
as a function of $\ell_1$, not their phases.

We emphasize here that the somewhat-challenging task of needing to determine
the (full-sky) $c_{\ell_1 m_1}$ discussed in this Appendix is largely orthogonal to the overall goal of
this paper of quantifying the effects of photometric variations. In practice,
the observers will have an opportunity to estimate the photometric calibration
variations across the sky in real space using a number of methods (e.g.\ using
theoretical estimates, or measurements across part of the sky as we did in our
examples). Converting from $c({\bf \hat{n}})$ to the full-sky $c_{\ell_1 m_1}$
is then just a purely mathematical exercise in which only an approximate
answer is required.

\section{Tabulation of coefficients}\label{app:tabulate}

In order to speed up calculations of the observed galaxy overdensity field
$\tlm$ and its correlations in harmonic space, we need to calculate a very
large number of geometrical quantities $R_{m_1 m_2 m}^{\ell_1 \ell_2 \ell}$;
see Eqs.~(\ref{eq:R})-(\ref{eq:U_in_terms_of_R}). Recall that each $(\ell, m)$
pair has $2\ell+1$ quantities for a fixed $\ell$, or approximately $\ellmax^2$
quantities for all $\ell$ up to some $\ellmax$. Calculating all of the $R$
coefficients would then naively involve $O(\ellmax^6)$ operations which, for
$\ellmax\simeq 1000$, would be $10^{18}$ calculations of the Wigner-3j
symbols! Calculating even a fraction of such a large number of coefficients is
clearly unfeasible.

We bring the computation required to a manageable size and speed as follows. We decide to tabulate the coefficients
$U$ for the minimum required number of quadruplets of its indices, and carry out the
summation in Eq.~(\ref{eq:U_in_terms_of_R}) for each $U$. Moreover, since our
goal is to concentrate on the large-angle (low-$\ell_1$) systematics, which limits
the maximum multipole {\it of the systematics}, $\ell_1$ (see
Eq.~(\ref{eq:U_in_terms_of_R})), we choose $\ell_1\leq \ellcalibmax$ with
$\ellcalibmax=20$. Therefore, the total number of evaluations of the
Wigner-3j symbols will be approximately equal to the number of $(\ell_1,
m_1)$ pairs times the number of the $U$ coefficients; the former quantity is
\begin{equation}
N_{\ell_1, m_1} = \sum_{\ell_1=0}^{\ellcalibmax}(2\ell_1 + 1) 
= (\ellcalibmax+1)^2,
\end{equation}
while the latter quantity we now evaluate.

At a fixed multipole of the {\it true} density field $\ell_2$ and a
calibration error at some multipole $\ell_1$, only the {\it observed}
multipoles in the range ${\rm max}\{(\ell_2-\ell_1), 0\}\leq \ell\leq
\ell_2+\ell_1$ are affected; this is well known feature of the coupling of
angular momenta in e.g.\ quantum mechanics. Therefore, for fixed values of
$\ell$ and $m$, the total number of $(\ell_2, m_2)$ pairs that can possibly
lead to nonzero $U$ coefficients is naively $N_{\ell_2, m_2}=
\sum_{\ell-\ellcalibmax}^{\ell+\ellcalibmax}(2\ell_2 + 1)$.  However, we must
also remember of a selection rule\footnote{The full list of selection rules
  for nonzero Wigner-3j symbols is: ${\rm max}[|\ell_i-\ell_j|, 0]\leq
  \ell_k\leq \ell_i+\ell_j$ and $|m_i|\leq \ell_i$ where $\{i, j, k\}$
  correspond to permutations of subscripts 1, 2, and no subscript. Moreover,
  $m_1+m_2+m=0$, and when $m_1=m_2=m=0$ then $\ell_1+\ell_2+\ell$ must be
  even.} that $m_1+m_2=m$, which means that for a fixed observed-field
coefficient $m$ and calibration-field coefficient $m_1$, only {\it one}
coefficient $m_2$ (instead of $2\ell_2+1$ of them) survives.  Therefore, for a
fixed $\ell$, the {\it total} number of $U$ coefficients to tabulate (e.g.\ for use
in Eq.~(\ref{eq:Tl})) is of order
\begin{eqnarray}
N_{m, \ell_2, m_2, \ell_1, m_1} 
&\simeq & (2\ell+1) (2\ellcalibmax + 1)N_{\ell_1, m_1}\nonumber\\[0.2cm]
&\simeq & 4\,\ell\,\ellcalibmax^3 
\end{eqnarray}
where the factors of $2\ell+1$ and $2\ellcalibmax+1$ refer to the number of
$m$ coefficients and $(\ell_2, m_2)$ pairs, respectively.
So for a typical calculation with $\ellcalibmax \simeq  20$, the number of
evaluations for a single multipole $\ell\sim 250$ is of order $10^7$; for of
order 30 values of $\ell$, including the densely spaced coverage of lowest
observable multipoles where the effects of photometric variation are the
largest, this is still less than $O(10^9)$ evaluations of the Wigner-3j
symbols, which is entirely feasible and takes of order a minute on a single 16-core
 desktop computer.
  
The last detail is actual computation of the Wigner-3j symbols: we use the
{\tt Gnu Scientific Library} (GSL) routines at $\ell\leq 50$. At higher
multipoles these break down, and we use a special-purpose
routine\footnote{\url{http://www.gnu.org/software/gsl/}.}  that uses
approximate expressions for the Wigner coefficients. These more accurate
expressions hold provided $\ell, \ell_2\gg \ell_1$, but given that $\ell>50$
in the regime we are using this approximation and the fact that the multipole
that corresponds to the calibration error is no bigger than
$\ell_1\leq\ellcalibmax=20$, this condition always holds.

\bibliographystyle{apsrev4-1}
\bibliography{de_ng}

\end{document}